\documentclass[useAMS,fleqn,usenatbib]{mn2e} 
\usepackage[T1]{fontenc}
\usepackage{ae,aecompl}
\usepackage{graphicx}    % Including figure files
\usepackage{currfile}
\title[WD pollution by steeply infalling bodies]           
{Deposition of steeply infalling debris around white dwarf stars}

\author[Brown, Veras \& G\"{a}nsicke]{                                                 
John C. Brown$^{1}$\thanks{E-mail: john.brown@glasgow.ac.uk},
Dimitri Veras$^{2}$,                      
Boris T. G\"{a}nsicke$^{2}$                                                     
\\                                                                              
$^{1}$School of Physics and Astronomy, University of Glasgow, Glasgow G12 8QQ, UK
\\
$^{2}$Department of Physics, University of Warwick, Coventry CV4 7AL, UK}

\pubyear{2016}

\begin{document}                                                                  
\pagerange{\pageref{firstpage}--\pageref{lastpage}}                               
\maketitle

\begin{abstract} 
High-metallicity pollution is common in white dwarf (WD) stars hosting remnant planetary systems. However, they rarely have detectable debris accretion discs, possibly because much of the influx is fast steeply-infalling debris in star-grazing orbits, producing a more tenuous signature than a slowly accreting disk. Processes governing such deposition between the Roche radius and photosphere have so far received little attention and we model them here analytically by extending recent work on sun-grazing comets to WD systems. We find that the evolution of cm-to-km size ($a_0$) infallers most strongly depends on two combinations of parameters, which effectively measure sublimation rate and binding strength. We then provide an algorithm to determine the fate of infallers for any WD, and apply the algorithm to four limiting combinations of hot versus cool (young/old) WDs with snowy (weak, volatile) versus rocky (strong, refractory) infallers. We find: (i) Total sublimation above the photosphere befalls all small infallers across the entire WD temperature $(T_{\rm {WD}})$ range, the threshold size rising with $T_{\rm {WD}}$ and 100$\times$ larger for rock than snow. (ii) All very large objects fragment tidally regardless of $T_{\rm {WD}}$: for rock, $a_0 \succeq 10^5$ cm; for snow, $a_0 \succeq 10^3-3\times10^4$ cm across all WD cooling ages. (iii) A considerable range of $a_0$ avoids fragmentation and total sublimation, yielding impacts or grazes with cold WDs. This range narrows rapidly with increasing $T_{\rm {WD}}$, especially for snowy bodies. Finally, we discuss briefly how the various forms of deposited debris may finally reach the photosphere surface itself.
\end{abstract}

%The current file is: \currfilename 


\label{firstpage} 
\section{Introduction}

The stratification of white dwarf (WD) atmospheres by atomic weight provides a tabula rasa upon which any deposited contaminants conspicuously stand out.  Abundant contaminants, in the form of heavy metals, have now been observed in one-quarter to one-half of all WDs \citep{zucetal2003,zucetal2010,koeetal2014}.  These metals cannot represent relics from stellar evolution because their diffusion (sinking) timescales are orders of magnitude shorter \citep{paqetal1986,wyaetal2014} than the age of the WDs (the {\it cooling time}).  The metals also cannot have predominantly arisen from the interstellar medium, which is too rarefied and hydrogen-rich \citep{aanetal1993,frietal2004,jura2006,kilred2007,faretal2010}.

Instead, the metals must originate from planetary system remnants \citep{gaeetal2012,juryou2014,xuetal2014,faretal2016,melduf2016}. This exciting development has been bolstered by strong evidence of at least one asteroid disintegrating in real time around a white dwarf \citep{vanetal2015,aloetal2016,gaeetal2016,garetal2016,rapetal2016,redetal2016,veretal2016c,xuetal2016,zhoetal2016,guretal2017} as well as nearly 40 dusty and gaseous discs orbiting within a distance of about one Solar radius \citep{zucbec1987,gaeetal2006,gaeetal2008,faretal2009,wiletal2014,baretal2016,denetal2016,farihi2016,manetal2016a,manetal2016b}. However, of the $\sim 1000$ metal polluted white dwarfs known, these 40 harbouring discs represent only a few per cent.

Hence a key question is how such white dwarfs without detectable discs become polluted? This pressing question has received little theoretical attention \citep{veras2016a} but it would seem likely either that the slowly infalling flat dense disk is too tenuous for detection or that the infall is substantially attributable to high-speed steep infall of tenuous matter in near-parabolic orbits of periastron distances near the stellar radius, and possibly is fairly isotropic. Simulations have shown that both grazing encounters \citep{musetal2014,vergae2015,hampor2016,petmun2016,veretal2016a,veras2016b} and even direct stellar impacts \citep{veretal2013,veretal2016c,veretal2016d} should occur. These encounters and potential impacts can include comets \citep{alcetal1986,veretal2014b,stoetal2015}, asteroids \citep{bonetal2011,debetal2012,frehan2014,antver2016} or even small moons \citep{payetal2016a,payetal2016b} though, as we show in this paper, these will become tidally fragmented nearer the star. Icy bodies like minor planets could easily retain internal water during the giant branch stages of evolution \citep{jurxu2010,jurxu2012,malper2016} so continued consideration of such bodies and even weaker snowy ones like comets is important.  The simulations in none of the above papers, however, described the physical details of the encounters, and just a handful of studies have considered physical aspects of the problem of destruction of bodies undergoing near-direct infall toward WD stars \citep{alcetal1986,beasok2015,stoetal2015}. 

Here we approach the problem of near-direct infall from a different perspective, by building on the analysis of the destruction regimes of steeply infalling solar comets by \cite{broetal2011} and \cite{broetal2015}. They consider both sublimation by starlight (at $r$ near $R_{\sun}$) and bow-shock ablation and ram pressure effects. Here, we extend their analysis to:  (i) the much larger range of parameters involved when one includes hard rocky infalling bodies; (ii) the very different stellar parameters of WD stars; and (iii) the tidal fragmentation regime (which \citealt*{broetal2011} and \citealt*{broetal2015} mostly ignored) because WD surface gravity is much stronger than solar gravity. Our aim is to determine (analytically and numerically) for what parameters the destruction of bodies of different sizes and properties is dominated by different processes, and the implications for WD pollution.

The primary goals of this paper are to: (i) elucidate the physics of steep infall of mall bodies toward WDs; and (ii) develop a useful algorithm for determining  the outcomes. In the following two sections we discuss the meaning, importance, range and uncertainty of values of all relevant WD stars (Section 2) and small body (Section 3) parameters. In Section 4  we describe the various relevant debris destruction  processes -- sublimation, fragmentation and grazing/impacts, and in Section 5 we  analyse which process(es) dominate in what spatial regime as a function of the parameters of the WD star and infalling object. We defer to Section 6 a discussion of the physics of debris impacting the photosphere directly. Then, in Section 7, we summarise our conclusion (7.1) and discuss briefly issues that need attention as to how debris (other than direct impactors) can finally reach the WD surface for the cases of sublimated matter, fragmented matter (where tides exceed strength) and steep star-grazers (near-misses) orbiting the WD. 

In order to aid the reader, we have summarized the meaning and location of the most important variables in Tables 3-4. 

\section{White Dwarf (WD) star properties}
\subsection{WD masses and radii} 
It is well known that WD stars occupy only a narrow range of masses $M_{\rm WD}$ because remnant WD masses much above 1 $M_\odot$ need rather large (so rare) progenitor masses while WDs of much below 1 $M_\odot$ are not reached by evolution within the current age of the universe.  Here we will consider WDs in the mass range $0.4-0.8M_\odot$ and in many of our results we use a mean value of $0.6M_\odot$ \citep{lieetal2005,faletal2010,treetal2016}.
The range of WD radii $R_{\rm WD}$ is even smaller because of the form of the mass-radius relationship \citep{hamsal1961} set mainly by the the hydrostatic balance of gravity and electron degeneracy pressure. For this we recognise that the relation is approximately independent of temperature over a wide range \citep{panetal2000}\footnote{More precise, but analytically less tractable relations, can be found in equations (4) and (5) of \cite{veretal2014a}.} and use the approximation.

\begin{equation}
R_{\star} =\gamma R_{\odot} \left[\frac{M_{\star}}{M_{\odot}}\right]^{-1/3}
\label{WDRoM}
\end{equation}
where 

\begin{equation}
\gamma \simeq 10^{-2}
\label{defgamma}
\end{equation}

\subsection{WD ages and effective temperatures}
WDs form with very high temperatures and cool at first very fast, but with rapidly decreasing rates by blackbody radiation. Their effective surface temperatures $T_{\rm WD}$ are an easily observed quantity, physically fixed by the cooling age $\tau_{\rm WDcool}$ of the star.  The relationship between $\tau_{\rm WDcool}$  and  $T_{\rm WD}$ also weakly involves  $M_{\rm WD}$ and has been studied in detail by \cite{mestel1952}, \cite{danmaz1990}, \cite{beretal1995}, and \cite{fonetal2001}, but the following is an adequate rough approximation  here.

\begin{equation} 
\tau_{\rm WDcool} \approx 700 {\rm Myr}\left(\frac{T_{\rm WD}}{10^4 \, {\rm K}}\right)^{-b}
\label{TWDagerelation}
\end{equation}
where $b\sim 4.5$ 

\subsection{Other WD quantities defined by mass and temperature}
The quantities ($M_{\rm WD}, T_{\rm WD}$) are sufficient to define the following quantities which also arise in our modelling.
The WD bolometric luminosity and radiation flux  at asterocentric distance $r$ are (with $x=r/R_{\star}$)
\begin{equation} 
F_{\rm rad}(r) = \frac{L_{\star}}{4\pi r^2} =\left[\frac{R_{\star}}{r}\right]^2 \sigma T_{\star}^4=\frac{F_{\star}}{x^2}
\label{Frad}
\end{equation}
while the  WD surface gravity is 
\begin{equation} 
g_{\star} = \frac{GM_{\star}}{R^2_{\star}}
\label{gravity}
\end{equation}
where $\sigma$ and $G$ are the Stefan-Boltzmann and Universal Gravitation constants. Likewise the surface escape speed $v_{\star}= (2GM_{\star}/R_{\star})^{1/2}$ which, for the WD case, becomes $v_{\star}=v_{\sun} \gamma^{-1/2} (M_{\star}/M_{\sun})^{2/3}$.
\subsection{WD atmospheres}
In the case of (grazing) infallers which are both large enough in size and small enough in periastron distance $q$ to reach the dense inner layers of the atmosphere, fluid interactions take over from radiation and tides and the density structure $\rho_{\rm  a}(r)$ of the atmosphere determines the destruction depth. For this case, we will use a locally exponential model $\rho_{\rm  a}(r)=\rho_{\rm  a0}\exp(-z/H)$  of constant scale height $H$  with $z=r$~$-R_{\rm WD}$. We will take the hydrostatic scale height to be that for ionised hydrogen (atomic mass $m_p$) at the WD effective temperature, viz

\begin{equation} 
H = \frac{  2kT_{\rm WD}} {m_p g_{\rm WD}}
\label{scaleheight}
\end{equation}

\noindent{}while the reference surface density  can be fitted to more realistic WD model atmosphere results such as those of \cite{treetal2011,treetal2013,treetal2015}.

For generality below, we will first derive debris destruction equations for stars of any ($M_{\star}, R_{\star}, T_{\star}$) including the solar case  ($M_{\sun}, R_{\sun}, T_{\sun}$) and then from them derive those for the WD case as a function of ($M_{\rm WD},T_{\rm WD}$) using the mass-radius relationship (\ref{WDRoM}). We do not concern ourselves here with variations in properties among the different classes of WDs because the large uncertainty and range in the infalling bodies (especially their strength) is much larger and dominates the uncertainties in our results.

\section{Properties of infalling bodies}
\subsection{Introduction}
In this section we will mainly be defining terminology and discussing typical values of properties of individual infalling bodies. We adopt homogeneous mean values through the body volume as reasonable - i.e. we consider the bodies to have individual {\textit{integrity}} in the way one would normally think of an asteroid, rock, pebble or hard-packed dirty snowball. However we recognise that many debris objects have only limited integrity, especially when it comes to strength, such as loose ice/snow/dust/rock conglomerates and inhomogeneous rocks containing cavities, cracks etc. -- or un-cemented rubble-piles or sand-heaps held together with almost exclusively self-gravity. For these it is essential to recognise the internal inhomogeneity of parameter values and its consequences. For example, for  a body made almost  entirely of hard rock, but permeated by cracks or surfaces of weakness, the local {\textit {strength}} of the rock material itself against strains is far higher than the effective strength of the body as a whole. Related is the ease or difficulty of pulling it apart into smaller stronger bodies of greater integrity -- i.e strength exceeding self-gravity. This important distinction will arise especially in Sections 3.5 and 4.3.

\subsection{Nucleus shape and size}
 We know from direct imaging, and from light curve data, that cometary nuclei and asteroids are of diverse, irregular and distinctly aspherical shapes. For our modelling purposes, therefore we characterise their linear size  by a single mean dimension $a$, their volume as $a^3$, and their direction-averaged cross-sectional area as $a^2$. If the shape were actually spherical with radius $a_{\rm s}$ and we chose $a=1.65a_{\rm s}$ then our expressions for the volume and the cross section would differ from the true values only by $\sim 10\%$. We define the initial size $a_0 \equiv a(r\rightarrow \infty)$ where $a(r)$ is its value at asterocentric distance $r$.

\subsection{Density $\rho$  and mass $M(r)$} 
We approximate the mass density $\rho$ -- and other intrinsic properties (e.g. $S$, $\mathcal{L}$, see below) -- of the infalling body as being uniform throughout its volume. Then the body's constant density $\rho$, and its evolving mass $M(r)$ and size $a(r)$ at $r$ are related -- provided it does not change shape or fragment --  by

\begin{equation}
M(r) = \rho a^3(r) 
\label{Mass(r)}
\end{equation}

The mean value widely used for the density of cometary objects is around half that of water (1 g/cm$^3$) but values vary somewhat between objects and estimates. They are thought to be comprised of a porous mix of ices, dust and rubble (\textit{dirty snowball}) which in this paper, for brevity, we will loosely term \textit{snow}. Here we allow for this variation by using a fiducial value $\rho_{\rm snow}=0.5$ g/cm$^3$ and writing the actual $\rho=\rho_{\rm snow}\times[\rho/\rho_{\rm snow}]$  when dealing with comet-like material with the dimensionless factor in square brackets selectable in a range of say 0.3 - 3.

Solid bodies like asteroids and pebbles (including solid ice) are denser, with $\rho$ in the range from around 1 g/cm$^3$ for solid ices to $\sim$ 10 g/cm$^3$ for bodies rich in iron ($\rho \simeq 8$ g/cm$^3$) and heavier materials.  We therefore define a fiducial value  used in numerical expressions of $\rho_{\rm rock}=3$ g/cm$^3$ and writing  $\rho=\rho_{\rm rock}\times[\rho/\rho_{\rm rock}]$  when dealing with rocky material with the with the dimensionless factor in square brackets again selectable in a range of say 0.3 - 3.

\subsection{Latent heat $\cal L$ }
The intrinsic parameters of an infalling body which mainly determine its rate of mass loss per unit area for a specified heating flux per unit area (see Section 4) are its density $\rho$ and latent heat ${\cal L}$ of sublimation/ablation. The relevant values of ${\cal L}$ for a star-grazing ice-conglomerate snowy mix for the regimes of intense heating where all components are vaporised was taken by \cite{broetal2011} (in their Section 2.2.2) to be the density-weighted mean over all mass components, including volatile and refractory ones,  which for a typical snowy cometary nucleus is  $\mathcal{L}_{\rm snow}\approx \mathcal{L}_{\rm }= 2.6\times10^{10}$~erg/g.  For solid rocky materials $\mathcal{L}$ is a few times higher: e.g. \cite{chyetal1993} adopted $2.3\times 10^{10}$ erg/g for comets, $8\times 10^{10}$~erg/g for stony/iron bodies and $10^{11}$ erg/g for solid iron. 

To allow for this range of values in both the snowy and rocky regimes we proceed similarly to what we did in subsection 3.2 for densities and write  ${\mathcal L} = {\mathcal L}_{\rm snow} \times[ {\mathcal L}/ {\mathcal L}_{\rm snow}]$ for snowy objects and  ${\mathcal L} = {\mathcal L}_{\rm rock} \times [{\mathcal L}/ {\mathcal L}_{\rm rock}]$ for rocky objects with fiducial values  in numerical expressions of ${\mathcal L}_{\rm snow}$~$=2.6\times10^{10}$ erg/g and ${\mathcal L}_{\rm rock}=8\times10^{10}$ erg/g the square-bracket factors again selectable in a range of say 0.3 - 3.

Note that some of our equations below involve the product $\rho{\mathcal L}$ which, for the above fiducial values, is about 20 times larger for rock than for snow. This in itself can be expected to yield very different behaviours of these two types of debris, but the difference in strength $S$ is even more dramatic as we now see.

\subsection{Strength $S$} 
In addition to sources of heat driving mass loss,  infalling bodies experience disruptive forces which can contribute to their dissipation  both directly and by accelerating their mass loss. These forces include: (i) the radial \textit{tensional} tidal force (gravity gradient) of the WD star; (ii) the associated azimuthal \textit{shear} force (from the orbital speed  gradient; \citealt*{davidsson1999,davidsson2001}); (iii) for bodies which enter the dense inner atmosphere of the WD, (\citealt*{broetal2011,broetal2015}) the \textit{compressional} force from the gradient of the ram pressure of the interaction with the dense atmosphere, and the lift force in the case of very shallow angle incidence; (iv) stellar radiation pressure, which can be important compared to gravity for dust particles. For the larger primary infalling bodies considered here ($a_0 >1$ cm), this resulting force can be neglected, especially in the very strong gravity of WDs; and possibly (v) the pressure gradient force arising from very intense sublimative mass outflow \citep{sekkra2015}. We will, however, return to the matter of radiation pressure when we consider briefly in Section 7 the final stage of arrival at the star of the much smaller debris (dust and atoms) created by sublimation and fragmentation.

The effects of these forces on the body depends on the ability of its material to resist them, i.e its relevant \textit{strengths: tensile,  shear, compressive}. Across the diverse infalling material properties, these strengths $S$ are the most wide-ranging in value. The different \textit{types} of strength can also differ considerably from each other for a single material but less so than the variation of a specific strength (e.g. tensile) between materials so here we will use solely tensile $S$  values as a starting point.

Consider for example, the mean local tensile strength $S$ which  (together with self-gravity) resists the disruptive  tidal gravity gradient.  This can be as high as a few times $10^{10}$ dyne/cm$^2$ for uniformly hard rocks (like granite). For comparison, the mean tensile strength value for the loose ice-conglomerate of some cometary nucleus material has been estimated from modelling and lab measurements to be  possibly as low as $10^3$ dyne/cm$^2$ (see e.g. \citealt*{greenetal1995}, \citealt*{gunandb2016} and references therein). A range of $10^3-10^7$ dyne/cm$^2$ has been reported for the localised surface  strengths of Comet 67P/Churyumov-Gerasimenko by \cite{bieetal2015} from two of the bounces  of ESA's Rosetta Philae lander but the strength relevant to bouncing is compressive rather than tensile, the latter being much smaller for loose materials. However, even a tensile strength equal to the smallest of all these $S$ values would be large enough to exceed self-gravity as the main adhesive force (see also Section 4.3 and Table 1) except for {\textit{very}} large bodies. However, the effective strength of many bodies to resist globally disruptive forces is often set by cracks and flaws which reduce the overall effective $S$ to near zero,  with ``zero-strength'' ``rubble piles'' of icy boulders or even sand heaps held together solely by self-gravity. A good example is that invoked to explain the ready breakup of some sungrazing comets and of Shoemaker-Levy 9 by Jupiter \citep{aspben1994}. We denote this class of body (self-gravity exceeding strength globally) as ``loose''.

In Section  4.3 we will argue that the high tidal gravity gradients around WD stars are so large that, inside the classical Roche limit -- which is far ($x \sim 100$) from the WD --  they act to pull apart  bodies of all sizes but only down to the limit of their constituent parts in which internal strength exceeds self-gravity. In other words bodies break up into parts bounded by their low-strength internal fault surfaces but no further at that stage. The infalling assembly of this processed debris will now comprise objects of a wide range of sizes but each of much higher integrity (and tensile strength) than its parent. In this region strength rather than self-gravity becomes the opponent of tidal fragmentation, with sublimative mass loss also limiting fragmentation to large bodies only (See Section 5).  In this inner sublimation region we will consider two broad classes of infalling matter strength: ``weak'' (like comet nucleus matter) and ``strong'' (rock) once again using the format $S=S_{\rm snow}\times[S/S_{\rm snow}]$ and  $S=S_{\rm rock}\times[S/S_{\rm rock}]$  for snowy and rocky bodies respectively, with fiducial values in numerical expressions of  $S_{\rm snow}=10^4$ and $S_{\rm rock}=10^{10}$ dyne/cm$^2$ respectively. The 6 order-of-magnitude range in $S$ values across debris types is a major factor in our findings, although an even wider range can be included in our modelling equations simply by adjusting $S/S_{\rm rock}$ or $S/S_{\rm snow}$ appropriately.

\begin{table*}
 \centering
  \caption{Our fiducial intrinsic parameter values for infalling objects.
}
  \begin{tabular}{@{}cccc@{}}
 \multicolumn{4}{c}{}   \\
  \hline
Type & Density $\rho$ &  Latent Heat $\cal{L}$ & Tensile Strength $S$  \\
     & (g/cm$^3$)  &  (erg/g)  &  (dyne/cm$^2$) \\
  \hline
  \hline
Rock & 3.0   & $ 8.0 \times 10^{10}$  & $10^{10}$    \\
Snow & 0.5   & $ 2.6 \times 10^{10}$  & $10^4$     \\
  \hline
\end{tabular}
\end{table*}

\subsection{Orbital geometry and speed}
\cite{broetal2011} and \cite{broetal2015} argued that, except in the very final stages of interaction with the deep atmosphere (when reached), or possibly in cases of rapid fragmentation \citep{sekkra2015}, the centre of mass of an infalling solid  body pretty much follows the locus and velocity of a Keplerian parabolic orbit about the central star.  \cite{broetal2011} and \cite{broetal2015} considered the general case of a parabola with arbitrary periastron distance $q$ (which also defines the stellar surface entry angle for cases where $q<R_{\star}$). 

In this paper we  are concerned with the behaviour of infalling material at the opposite extreme from the slow inflow of accretion disc matter as viscosity redistributes angular momentum -- namely the case of infallers having orbital eccentricities near unity and very small perihelion distances $q$ close to or less than $R_{\rm WD}$ (with very low angular momentum). We will loosely term all of  these {\textit {impactors}} but intend to include both those that could actually impact the atmosphere (star-plungers or -divers) and those that would have near-miss fly-bys (star-grazers) {{\textit {if}} they are not fully sublimated or fragmented before getting that close. For such objects, along most of their path inside the Roche limit ($x\sim 100$) down to $x$ of a few, the trajectory and velocity are nearly identical to those of a  linear parabolic orbit ($q=0$). Through the next two sections we consider them as such to simplify the mathematical treatment of the sublimation/fragmentation processes (see also \citealt*{broetal2011}) who treat exactly the case of general $q$. In Section 6  we allow for the effect of deviations from the linear parabolic trajectory in addressing the behaviour (entry angle, etc.) of material actually impacting the WD surface and in Section 7 how near-miss star-grazing material might find its way onto the surface.

In the regime ($r\gg R_{\rm WD}$) of the linear parabola approximation, the only component of the velocity vector is the radial speed $v(r)=v_r(r) =\dot r = dr/dt$ with the property that

\begin{equation}
\frac{dv}{dt}\equiv\dot v = v\frac{dv}{dr} = -\frac{GM_*}{r^2}
\label{acceleration}
\end{equation}

\noindent{}yielding (for $v(r\rightarrow\infty) = 0$) the usual solution 

\begin{equation}
v(r)=
\left(
\frac{2GM_{\star}}{r}
\right)^{1/2} 
= 
v_{\star}
\left(
\frac{R_{\star}}
{r}
\right)^{1/2}
.
\label{vofr}
\end{equation}

\section{Processes of destruction of infalling bodies}
\subsection{Overview}
As noted above, destruction of infalling bodies occurs by a combination of: (a) sublimative mass loss by an energy flux $F$ of starlight that is sufficiently large to raise the bodies above the vaporisation temperature of at least some, and eventually all, of their components; (b) fragmentation due to the stellar tidal {or possibly internal pressure} forces exceeding the internal strength and self-gravity of the body; and (c) frictional ablative mass loss and ram pressure pancaking and deceleration effects in the dense low atmosphere. 

The importance of these  various processes all decline with distance $r$ but at at differing rates. Stellar radiation flux (sublimation) declines as $\propto 1/r^2$ while tidal forces decline faster as $\propto 1/r^3$ and effectively cut-off at a finite distance when self-gravity and body strength offset them. Finally, atmospheric frictional ram pressure effects vary as $\sim \rho_{\rm atm}(r)$ where the atmospheric mass density  $\rho_{\rm atm}(r) \propto \exp{(-(r-R_{\star})/H})$ near $r=R_{\star}$ with scale height $H\ll R_{\star}$. Consequently these only become  important within a  few $H$ of $r=R_{\rm WD}$. Disruptive internal pressure is also only  important if very high mass loss rates arise near the star (e.g. \citealt*{sekkra2015}).

We then argue (in Section 4.3.2),  by analogy  with observations of  fragmentation of some sun-grazers and of SL-9 by Jupiter and the modelling of the latter by \citet{aspben1996}, that the initial process is tidal disruption of large very loose structures into smaller more robust components in which strength everywhere exceeds self gravity.

Thereafter as infall progresses from large $r$, radiative sublimation dominates until fragmentation sets in (if at all) or bow-shock ablation/deceleration takes over near the photosphere. Below we therefore first model (Section 4.2) the infall evolution of $a(r)$ and $M(r)$ assuming that only sublimation is active. Then in Section 4.3 we discuss the basics of tidal fragmentation and (see also \citealt*{beasok2015}) include the effects  of strength as well as of self-gravity. The latter is often ignored but in fact proves to be dominant even for \textit{quite} weak snowy cometary material as we show below.  Then in Section 5  we discuss the interplay of sublimation and tidal fragmentation  as a function of original infaller size.  We determine the spatial and parametric ranges for which tidal fragmentation may dominate over sublimation, drawing conclusions relevant to the WD debris infall problem. We defer to Section 6 treatment of the details of destruction of objects ({\textit {impactors}}) which are large/strong enough to enter the high density gas layers low in the atmosphere where destructive hydrodynamic effects abruptly take over from sublimation and tidal forces. In Section 7.1 we discuss briefly the issues involved in whether and how sublimated atomic/molecular matter and fragmented pebbly/dusty debris can reach the stellar surface and  in 7.2 what happens to original infallers or to their fragmented pieces. These pieces would be close to but not quite in the regime of direct infall, but rather they have orbits grazing close by the photosphere -- specifically we address how they may shed enough of their angular momentum to reach the photosphere.

One other destructive force proposed  (in the solar comet context)  as sometimes important in infalling debris deposition is \citep{steetal2015} the ram pressure $P_{\rm ramsub}$  of sublimating mass outflow when that outflow is high (near the star). Based on the rates found by \cite{broetal2011} or the equations of the next subsection, one finds that for comets quite near the sun, $P_{\rm ramsub}$ can exceed the low strength $S$ of cometary, but not that of rocky, material. However, \cite{gunetal2012} argued that if the outflow is symmetric enough, the inward reaction force to the outflow pressure $P_{\rm ramsub}$ can \textit{oppose} fragmentation. On the other hand, \cite{sekkra2015} have invoked an energetic exothermal process in ice crystal formation to explain the sudden fragmentation of Comet  C/2012 S1 Ison  while still well outside the Roche Lobe. 
Such processes may have to be considered for some stellar infall but we omit them here as their importance is not yet widely agreed upon by the community.

\subsection{Sublimation in starlight}

By taking the infalling body to have near zero albedo, the heating power of starlight entering the infalling body  is $a^2 F_{\rm rad}$, with the stellar radiation flux $F_{\rm rad}$ is given by  Equation (\ref{Frad})\footnote{We neglect here the correction factor $\sim 1-2$ that is strictly required in Equation \ref{Frad} as $r$ comes close to $R$ because at that location the stellar radiation flux is not unidirectional, but rather arises from the large finite angular size stellar disc.}. We also neglect radiative cooling, assuming that the sublimation  occurs on a  timescale faster than that of radiative energy loss. This neglect is based on the fact that the specific energy $\mathcal{L}  \sim  2.6 \times 10^{10}$ erg/g $\sim$~0.01~eV/nucleon needed for sublimation of ice and rock is much smaller than that needed to heat it to the radiative equilibrium $T_{\rm eq}\sim T_{\star}/x^{1/2}$ viz $kT_{\rm eq}/m_{\rm p}\sim 0.6/x^{1/2}$ eV. An exception occurs at $x \gg 1$, where, in any case $T_{\rm eq}$ falls below the sublimation threshold temperature and little sublimation occurs. Then, for a body of density $\rho$ and latent heat $\cal L$, with mass $M(r)$, and size $a(r)$  ( $\rightarrow M_0,a_0$ as $r\rightarrow\infty$) the mass loss per unit radial distance is, using equations (\ref{acceleration}) and (\ref{vofr}),

\begin{eqnarray}
\frac{dM}{dr} &=& \frac{1}{v(r)}\frac{dM}{dt}=\frac{1}{v_{\star}}\left(\frac{r}{R_{\star}}\right)^{1/2}\frac{F_{\rm rad} a^2}{\mathcal{L}}
\nonumber\\
              &=&\frac{\sigma T_{\star}^4}{{\cal L} v_{\star}\rho^{2/3}}\left(\frac{R_{\star}}{r}\right)^{3/2}M^{2/3}
\label{dmdr}
\end{eqnarray}
By using $M_0=\rho a_0^3$, $M=\rho a^3$ and $v_{\star} = (2GM_{\star}/R_{\star})^{1/2}$, we obtain a solution for the variation with relative distance $x=r/R_{\star}$ of size and mass $a,M$ (with original incident values $a_0,M_0$)

\begin{equation}
a(x)= a_{\rm sub}(x)=a_0-\frac{A}{x^{1/2}}
\label{aofAandx}
\end{equation}

\begin{equation}
\frac{M(x)}{M_0}=\left(\frac{a(x)}{a_0}\right) ^3 =\left(1-\frac{A}{a_0x^{1/2}}\right)^3
\label{MofAandx}
\end{equation}

\begin{equation}
A = \frac{2R_{\star}\sigma T_{\star}^4}{3\rho{\mathcal L}v_{\star}}=\frac{2^{1/2}R_{\star}^{3/2}\sigma T_{\star}^4}{3\rho{\mathcal L}G^{1/2}M_{\star}^{1/2}}
\label{Adef}
\end{equation}
where the {\textit {sublimation parameter}} $A$ (cm)  is clearly the minimum initial size $a_0$ of object needed to just survive sublimation alone down to the photosphere ($r=R_{\star}, x=1$) and represents a crucial value in the debris deposition problem.

An important result of Equation (\ref{aofAandx}) is that in the absence of fragmentation the sublimative drop in size $a(x)$ at $x$ is $A(x)$ independent of $a_0$ and in particular the size $a(1)$  of an un-fragmented object reaching the photosphere is

\begin{equation}
a(1)=a_0-A
\label{aof1}
\end{equation}

For a general star characterised by $M_{\star},R_{\star},T_{\star}$ we can rewrite (\ref{Adef}) as 

\begin{equation}
A_{\star}= \left(\frac{2^{1/2}R_{\odot}^{3/2}\sigma T_{\odot}^4}{3\rho{\mathcal L}G^{1/2}M_{\odot}^{1/2}}\right)
\left[
\frac{
\left(R_{\star}/R_{\odot}\right)^{3/2}
\left(T_{\star}/T_{\odot}\right)^{4}
}
{
\left(M_{\star}/M_{\odot}\right)^{1/2}
}
\right]
\label{Astargen}
\end{equation}
For any WD star that is characterised by $M_{\rm WD},T_{\rm WD}$, by using the $R_{\rm WD}(M_{\rm WD})$ relationship (\ref{WDRoM}), we obtain the following expression:

\begin{equation}
A_{\rm WD}= \left(\frac{2^{1/2}R_{\odot}^{3/2}\sigma T_{\odot}^4}{3\rho{\mathcal L}G^{1/2}M_{\odot}^{1/2}}\right)\gamma^{3/2}
\left[
\frac{
\left(T_{\rm WD}/T_{\odot}\right)^{4}}
{
\left(M_{\rm WD}/M_{\odot}\right)
}
\right]
\label{AWD}
\end{equation}
This relation leads to the following numerical expressions for general stars and for WD stars.  Each group is given in two distinct forms: one expressed in terms of values of  $\rho, {\cal L}, S$ relative to our fiducial values for rock and the other relative to our fiducial values for snow.
In the WD cases we have replaced $\left(T_{\rm WD}/T_{\odot}\right)$ by $\left(T_{\rm WD}/10^4 \, {\rm K}\right)$ as the latter is more convenient to the WD community. These fiducial values are all given in Table 1. 

\begin{equation}
A_{\star}^{\rm rock} ({\rm cm}) =
\left(
\frac{2.0\times 10^2}
{(\rho/\rho_{\rm rock})({\mathcal L}/{\mathcal L}_{\rm rock})}
\right) 
\left[
\frac{
\left(\frac{R_{\star}}{R_{\odot}}\right)^{3/2}
\left(\frac{T_{\star}}{T_{\odot}}\right)^{4}
}
{
\left(\frac{M_{\star}}{M_{\odot}}\right)^{1/2}
}
\right]
\label{Astarrock}
\end{equation}

\begin{equation}
A_{\star}^{\rm snow}({\rm cm})= \left(\frac{3.7\times10^3}
{(\rho/\rho_{\rm snow})({\mathcal L}/{\mathcal L}_{\rm snow})}
\right)
\left[
\frac{
\left(\frac{R_{\star}}{R_{\odot}}\right)^{3/2}
\left(\frac{T_{\star}}{T_{\odot}}\right)^{4}
}
{
\left(\frac{M_{\star}}{M_{\odot}}\right)^{1/2}
}
\right]
\label{Astarsnow}
\end{equation}

\begin{equation}
A_{\rm WD}^{\rm rock}({\rm cm})= \left(\frac{1.8}   {(\rho/\rho_{\rm rock}) (\mathcal L/\mathcal L_{\rm rock})}
\right)
\left[
\frac{
\left(T_{\rm WD}/10^4 \, {\rm K}\right)^{4}}
{
\left(M_{\rm WD}/M_{\odot}\right)
}
\right]
\label{AWDRock}
\end{equation}

\begin{equation}
A_{\rm WD}^{\rm snow}({\rm cm})= \left(\frac{33}   {(\rho/\rho_{\rm snow}) (\mathcal L/\mathcal L_{\rm snow})}
\right)
\left[
\frac{
\left(T_{\rm WD}/10^4 \, {\rm K}\right)^{4}}
{
\left(M_{\rm WD}/M_{\odot}\right)
}
\right]
\label{AWDSnow}
\end{equation}
The second expression ($A_{\star}^{\rm snow}$) agrees numerically  with the results of the \cite{broetal2011} paper on solar comets for $M_{\star}=M_{\odot}, R_{\star}= R_{\odot}$ in the case of zero periastron. 

The above two expressions for $A(T)$ in our fiducial white dwarf cases are shown in Figure 1.

%%%%%%%%%%%%%%%%%%%%%%%%%%%%%%%%%% 
\begin{figure}
\centerline{
\includegraphics[width=8cm,height=6cm]{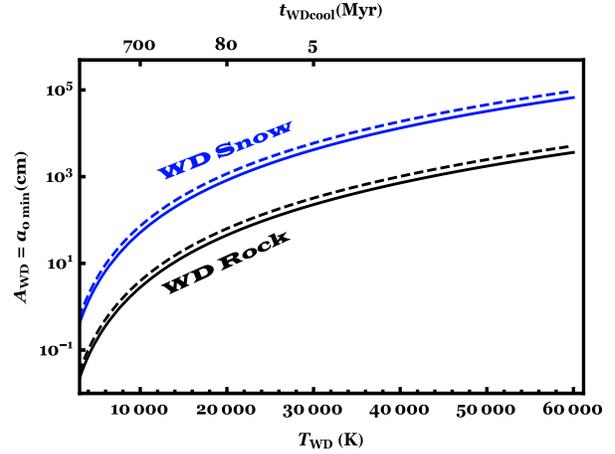}
}
\caption{
Plot of $A_{\rm WD}$ versus $(T_{\rm WD}$ in the range $3000< T_{\rm WD}({\rm K})<60000$ for rocky (lower pair of lines) and snowy (upper pair of lines) fiducial parameters and for masses $M_{\rm WD}=$~$0.4,0.8M_{\sun}$ (dashed and solid curves). Note that $A_{\rm WD}=a_{\rm 0min}$ the minimum incident size needed to survive sublimation alone (no fragmentation) down to the WD surface. The top axis shows the mean cooling time $t_{\rm WDcool}$ (for an average $M_{\rm WD}$ value) corresponding to the values of $(T_{\rm WD}$ along the bottom axis. $t_{\rm WDcool}(a_{\rm 0min})$ is thus the time after which objects of initial size $a_{\rm 0min}$ can just reach the surface without total sublimation.
}
\label{FigBig1}
\end{figure}
%%%%%%%%%%%%%%%%%%%%%%%%%%%%%%%%%% 

By comparing the value of $A$ given by equation (\ref{Astargen}) for the sun with that from equation (\ref{AWD}) for a WD of the same mass and temperature, we see that an object 1000 times smaller  can survive sublimation down to the surface of the WD compared to the size needed to reach the surface of the sun. This result is due to the fact that although the sublimating starlight has the same flux near the stellar surface in both cases, the effective time of exposure to that flux scales as the infall time $R_{\star}/v_{\star}\propto R_{\star}^{3/2}$ which is $\gamma^{3/2}=10^{-3}$ times smaller for a WD than for the sun. In order for the minimum incident $a_o=A$ value to allow an object to survive to the photosphere to be the same as for the sun, the WD of the same mass would have to be hotter than the sun by a factor $10^{3/4}$ or  $T_{\rm WD} \approx 32,600$ K. For smaller WD masses, the minimum size for survival increases, partly because of $R$. Hence, infall time is larger, but also because the luminosity $\propto R^2$, and so sublimative mass loss increases.

\subsection{Tidal fragmentation including material strength}
\subsubsection{General Case}

Neglecting internal pressure forces, the net disruptive force ($\mathcal{F}_{\rm tot}$) across a small infalling body (size $a$)  is the difference between the disruptive tidal force ($\mathcal{F}_{\rm T}$), and the sum of the binding self-gravity ($\mathcal{F}_{\rm G}$)  and tensile strength  ($\mathcal{F}_{\rm S}$) forces. (The following summary of forces is similar to that in \citealt*{beasok2015} within factors of order unity).

\begin{equation}
\mathcal{F}_{\rm tot} = +\left|\mathcal{F}_{\rm T}\right| -\left|\mathcal{F}_{\rm S}\right| - \left|\mathcal{F}_{\rm G} \right|
\label{forcesum}
\end{equation}

\noindent{}Standard approximations for each component are

\begin{eqnarray}
\mathcal{F}_{\rm T} &\approx& \frac{G M_{\star} M a}{2 r^3} 
,
\\
\mathcal{F}_{\rm G} &\approx& -\frac{G M^2}{a^2} 
,
\\
\mathcal{F}_{\rm S} &=& - S a^2
.
\label{forcevalues}
\end{eqnarray}
The condition for a body to remain intact is thus (replacing $M$ by $\rho a^3$ and setting $M_{\star}= 4\pi \rho_{\star} R_{\star}^3/3)$ that $a$ should satisfy
\begin{equation}
\frac{\left|\mathcal{F}_{\rm S} + \mathcal{F}_{\rm G}\right|}{\mathcal{F}_{\rm T}} =\frac{S/(G\rho^2a^2)+1}{M_*/(2\rho r^3)}=\frac{S/(G\rho^2a^2)+1}{(2\pi/3)(\rho_{\star}/\rho x^3)} \geq 1
\label{Nofragcond}
\end{equation}

\noindent{}which defines the maximum size $a_{\rm frag}$ of object that can avoid fragmentation at distance $r=xR_{\star}$ namely 

\begin{equation}
a(x) \le a_{\rm frag} (x)= 
\sqrt{ \frac{S/G\rho^2}{(2\pi/3)(\rho_{\star}/\rho x^3) -1}}
\label{amaxfragfull}
\end{equation} 

In this paper we will mainly be discussing the properties of this equation in the limit  where the strength $S$ term dominates over self-gravity -- see Section 4.3.3 -- but first we look at the opposite low $S$ limit of ``loose'' incoming material held together only by self-gravity, which was mentioned already in Sections 3.1 and 3.5.

\subsubsection{The loose (zero $S$) self-gravity dominated Roche-limit regime}

In the (loose)  limit  
\begin{equation}
\mathcal{F}_{\rm S}/ \mathcal{F}_{\rm G} \ll 1
\end{equation}

Equation (\ref{Nofragcond}) simplifies to the usual Roche limit form

\[
x>x_{\rm Roche}=\left(\frac{2\pi\rho_{\rm WD}}{3\rho}\right)^{1/3}
\]
\begin{equation}
\ \ \ \ \ \ \ \ \ \ \ \ \ \ \ =\frac{1}{\gamma}\left(\frac{2\pi\rho_{\sun}}{3\rho}\left(\frac{M_{\rm WD}}{M_{\sun}}\right)^2   \right) ^{1/3}
\label{Rochex}
\end{equation}
which means typically $r_{\rm Roche}=\gamma x_{\rm Roche} R_{\rm WD} \sim 100 R_{\rm WD} \sim 1R_\odot$. This result is expected because the gravity gradient of a $1M_\odot$ WD is the same as that of the sun at the same absolute distance, regardless of the size of the object concerned, so long as it is \textit{loose} with self-gravity dominating over strength. 

As already mentioned in Sections 3.1 and 3.5, many incident debris object are likely initially to be \textit{loose} in the sense of being aggregates of smaller more internally solid pieces, or permeated by cracks or other faults. The {\textit{global} strength of such objects against disruption of its loose components can be very small, far less than the \textit{internal}  strengths of the latter \citep{aspben1996}. We thus envisage the scenario that all debris approaching a WD within around 100 WD radii  will tend to be tidally fragmented into smaller component volumes containing matter of much greater integrity and with global strength much higher than that of the object as  a whole and than self gravity. Inward of this distance the effects of tides on the remnant objects will be dependent on the greater strength $S$ of their material, be it \textit{weak} or \textit{strong}, as well as on distance $x$, as we analyse in Section 4.3.3. This scenario is just what we observed for Shoemaker-Levy 9 as it approached Jupiter \citep{aspben1996}.  The above equations show that the strength $S$ needed in an object of size $a$ and density $\rho$ for strength to dominate over self gravity is only $S > G\rho^2a^2 = 175(a \ {\rm (km)})^2$ (dyne/cm$^2$) for the density of snow and $6300(a \ {\rm (km)})^2$ for the density of rock.  These $S$ values are very small except for large objects (km and up).

\subsubsection{The {weak and strong} $S$-dominated regime}
When we now consider Equations (24)  and (25)  for \textit{large} $S$, we recall that fragmentation onset is no longer solely  a function of the infaller distance $x$ and density as in the loose ({\textit {rubble pile}}) Roche limit case, but also of the infaller strength $S$ and size $a$. We find the condition on $a(x)$  to avoid fragmentation is 

\begin{equation}
a(x) < a_{\rm frag} (x)= 
\sqrt{ \frac{S/G\rho^2}{(2\pi/3)(\rho_{\star}/\rho x^3 -1)}}\approx Bx^{3/2}
\label{amaxfrag}
\end{equation} 
where 
\begin{equation}
B=\sqrt{ \frac{3S}{2\pi G\rho\rho_{\star}}}=\sqrt{ \frac{2SR^3_{\star}}{G\rho M_{\star}}}
\label{defB} 
\end{equation} 
The final expression amounts to neglecting self-gravity as opposed to strength and is a good approximation whenever $\rho_{\star}\gg\rho x^3$.  Later results show that for white dwarfs, this approximation is valid except for {\it very} large, weak and low-density infallers. 
%Fig. \ref{FigBig4}
Essentially it amounts to dropping the -1 from the denominator of Equation (\ref{amaxfrag}), which arose from the self-gravity term $\mathcal{F}_{\rm G}$. It emphasises the fact, not widely appreciated, that even for materials of low $S$ like snowy cometary nuclei, self gravity $\mathcal{F}_{\rm G}$ is unimportant compared to material strength $\mathcal{F}_{\rm S}$ in opposing tidal fragmentation except for quite large objects. This can be seen by examining the ratio 
 
\begin{eqnarray}
{\mathcal R}=\frac{\mathcal{F}_{\rm G}}{\mathcal{F}_{\rm S}} = \frac{G\rho^2a^2}{S} 
\label{AppEqn1a}
\\
{\mathcal R}_{\rm rock} =  0.6\frac{(\rho/3\ {\rm g \, cm}^{-3})^2(a/{\rm \, 1000 km})^2}{(S/10^{10}\ {\rm dyne \, cm}^{-2})} 
\label{AppEqn1b}
\\
{\mathcal R}_{\rm snow} = 0.017\frac{(\rho/0.5\ {\rm g \, cm}^{-3})^2(a/{\rm \, 1 km})^2}{(S/10^{4}\ {\rm dyne \, cm}^{-2})} 
\label{AppEqn1}
\end{eqnarray}

In Table 2 we show ${\mathcal R}_{\rm rock}$ and  ${\mathcal R}_{\rm snow}$  versus $a({\rm cm})$. 
For our fiducial cometary value of $S=10^4$ dyne/cm$^2$ (Table 2)  the size has to exceed about 1 km (mass around $10^{15}$ g, as in C/2011 W3 Lovejoy) for self-gravity to dominate over strength, while for  $S=10^3$ dyne/cm$^2$ the minimum size is around 100 m. For rocks with  our fiducial $S=10^{10}$ dyne/cm$^2$  the minimum is around 1000~km which is why only asteroids/dwarf planets larger  than this size tend toward sphericity (isotropic self-gravity defeats anisotropic rock strength). The relevance of the classical Roche tidal limit (based on self-gravity alone) is solely for the disassembling of aggregations of bodies which are very loosely bound or unbound (apart from by self-gravity) such as rubble- or sand-piles. It is not relevant to scales on which the constituent bodies (individual boulders, sandgrains, ice crystals etc) have integrity in the sense of significant internal strength with no weak fracture planes, as already mentioned in Sections 3.1 and 3.5.

\begin{table}
\caption{
Ratio of forces of self-gravity to 
internal strength.}
\label{TableAbbreviations}
\begin{tabular}{ccccc}
\hline
Type & $a$ (cm) & $\mathcal{R}$  \\
\hline
Rock & $10^2$ & $6.0 \times 10^{-13}$ \\
Rock & $10^3$ & $6.0 \times 10^{-11}$ \\
Rock & $10^4$ & $6.0 \times 10^{-9}$  \\
Rock & $10^5$ & $6.0 \times 10^{-7}$  \\
Rock & $10^6$ & $6.0 \times 10^{-5}$  \\
Snow & $10^2$ & $1.7 \times 10^{-8}$ \\
Snow & $10^3$ & $1.7 \times 10^{-6}$ \\
Snow & $10^4$ & $1.7 \times 10^{-4}$  \\
Snow & $10^5$ & $1.7 \times 10^{-2}$  \\
Snow & $10^6$ & $1.7$  \\
\hline
\end{tabular}
\end{table}%%%End of the table

$B$ (cm) is clearly a {\textit {binding size parameter}} measuring the threshold size which must be exceeded for the tidal force to overcome its strength and thus for fragmentation to set in. It is also obviously the minimum size of object which would fragment if placed directly at $x=1$. By considering Equation (\ref{amaxfrag}) we can also see that $B$ can be expressed in terms of the relative distance $x = (a_o/B)^{2/3}$ at which a body of initial size $a_o$ would start to fragment in the 
absence of any significant sublimative reduction in size (i.e. in the limit of very small $A$ due, for example, to very large ${\cal L} $ or low $T_{\rm WD}$). In reality, as we discuss in Section 5, one must consider the interplay of processes with sublimative decline of $a(x)$ allowing deeper infall before fragmentation.

Thus for a general star we can write 

\begin{equation}
B_{\star} = 
\sqrt{\frac{2SR^3_{\star}}{G\rho M_{\star}}}=\sqrt{\frac{2SR^3_\odot}{G\rho M_\odot}}\frac{(R_{\star}/R_\odot)^{3/2}}{(M_{\star}/M_\odot)^{1/2}}
\label{Bstar}
\end{equation}
By using the $R(M)$ relation for WDs, in their case it becomes

\begin{equation}
B_{\rm WD} =\gamma^{3/2} \sqrt{\frac{2R^3_\odot}{GM_\odot}}\left(\frac
{S}{\rho}\right)^{1/2}\frac{1}{(M_{\rm WD}/M_\odot)}
\label{BWD}
\end{equation}

\noindent{}The corresponding numerical values for rock and snow infallers to stars and to WDs are

\begin{equation}
B_{\star}^{{\rm rock}} ({\rm cm})
= {1.3\times 10^8} \left( \frac{S/S_{\rm rock}}{\rho/\rho_{\rm rock}} \right)^{1/2}
  \frac{(R_{\star}/R_\odot)^{3/2}}{(M_{\star}/M_\odot)^{1/2}}
\label{BstarRock}
\end{equation}

\begin{equation}
B_{\star}^{\rm snow} ({\rm cm})
= {3.2\times 10^5} \left( \frac{S/S_{\rm snow}}{\rho/\rho_{\rm snow}} \right)^{1/2}
  \frac{(R_{\star}/R_\odot)^{3/2}}{(M_{\star}/M_\odot)^{1/2}}
\label{BstarSnow}
\end{equation}

\begin{equation}
B_{\rm WD}^{{{\rm rock}}} ({\rm cm})
= {1.3\times 10^5} \left( \frac{S/S_{\rm rock}}{\rho/\rho_{\rm rock}} \right)^{1/2}
  \frac{1}{(M_{\rm WD}/M_\odot)}
\label{BWDRock}
\end{equation}

\begin{equation}
B_{\rm WD}^{{\rm snow}} ({\rm cm})
= {3.2\times 10^2} \left( \frac{S/S_{\rm snow}}{\rho/\rho_{\rm snow}} \right)^{1/2}
  \frac{1}{(M_{\rm WD}/M_\odot)}
\label{BWDSnow}
\end{equation}

It is evident from Equations (\ref{BstarRock}) - (\ref{BWDSnow}) that, while similar to the solar case for similar absolute $r$ values,  tidal fragmentation is far more important for WDs than for the sun for $r$ values nearing the stellar radius. The reason is because of the much larger stellar surface tidal force gradient ($\propto GM/R^3$ from equation \ref{defB}, so a factor of $10^6$ larger for a given mass). Though $B=a_{\rm frag}(x=1)$ is the largest size of object that could be placed directly at the photosphere without fragmenting, allowing for sublimative loss of size during infall, in the absence of fragmentation, corresponds to an infalling object of initial size $a_0= B+A$.

\subsection{Destruction by atmospheric impact}
As we will confirm in Section 5, regimes  exist where intact incident objects (or intact components of tidally fragmented loose incident objects) partially survive sublimation down to $x \sim 1$, where they undergo destruction by  {\textit {impact}}  with the stellar atmosphere or possibly disruptive processes in a grazing near-miss.  We defer detailed modelling of such cases to Section 6 after we have discussed the parameter regimes in which it, and also sublimative and fragmentational destruction, occur.

\section{Destruction parameter regimes }

\subsection{Basics}
In Sections 4.2 and 4.3 we have discussed separately how body size $a(x,a_0)$ declines with distance $xR_{\rm WD}$ due to sublimation alone (Equation \ref{aofAandx}) and how the maximum size $a_{\rm frag}$ which can survive tidal fragmentation alone declines with $x$. This is all for $x$ smaller than the boundary where tidal forces inside the Roche (self-gravity  dominated) limit have already processed very loosely bound infalling debris into smaller chunks of higher integrity in which internal strength far exceeds self gravity. So it is clear that there is interplay between all the processes (see Sections 6 and 7). Qualitatively it is apparent that the possible fates of a stellar infall object whose cohesion is from internal strength are as follows:
\begin{enumerate} 
\item Total {\textit {SUBLIMATION}} outside the photosphere without encountering its tidal fragmentation limit.

\item Here we lump together, under the heading {\textit {IMPACT}}, both (a) actual impacts ($q<R_{\rm WD}$) and (b) very close grazing encounters (say $R_{\rm WD}<q \ll  2R_{\rm WD}$) with the photosphere before complete  sublimation or reaching its fragmentation limit. Destruction in the actual impact case is by abrupt ablative mass loss and deceleration via bow-shock interaction with the dense stellar atmosphere, as detailed in Section 6. The final fate of matter undergoing grazing near misses is discussed in Section 7 along with  whether and how the products of fragmentation and of sublimation well above the photosphere eventually reach it.

\item  {\textit {FRAGMENTATION}} occurs before total sublimation or photospheric impact.  It is beyond the scope of this paper to analyse in detail the behaviour of such fragmenting bodies (see also \citealt*{aspben1996}) and here we mainly refer to their initial fragmentation point as their end point  as it signifies the demise of the original single bodies. However, we note that  the behaviour of any of the smaller objects resulting from the first fragmentation which retain integrity (in the sense of $S$  once again dominating over the tidal force) can be followed along the same lines as our treatment above until they either sublimate totally, impact the photosphere, or reach their own fragmentation limit and fragment again. We examine briefly below some special cases of this hierarchical fragmentation.  (Also,  in Section 7, we touch on other processes affecting post-fragmentation  evolution).

We can shed some light here on hierarchical fragmentation theory for cases where fractional sublimative decrease in size $a(x)$ between successive fragmentations is small. This approximation applies for $B\gg A$ -- for example to large enough $a_0, {\cal{L}}$ and/or low enough $T_{\rm {WD}}$. Then, by equation (\ref{amaxfrag}), an object of finite tensile strength $S$ and  size $a_1$ first reaches its tidal disruption limit at distance $x_1= (a_1/B)^{2/3}$ (if $a_1>B$ so that $x_1>1$). From that point onwards one can imagine two limiting behaviours. 

The first is a {\textit {marginally stable}} progression in which the mass and size of the infalling body are almost steadily diminished by tidal loss of small fragments such that $a(x)$ is kept just at the the limiting value $a(x)=Bx^{3/2}$ with the body mass $M(x)$ declining with $x$ according to $M(x)/M_1=\left[a(x)/a_1\right]^3 =(x/x_1)^{9/2}$ and arriving at the photosphere ($x=1$) with precisely the critical tidal disruption size $B$ there, the  rest of the mass having been transformed into small particles en route.

Secondly, at the other extreme, we can envisage each tidal disruption occurring at $x=x_1$  and subsequent critical points, if any,  to take the form of breakup into $j$ equal parts with $j\ge 2$. Then, inside $x_1$, the object initially comprises $j$ parts, each of  mass $M_2=M_1/j =\rho a_1^3/j$ and size $a_2=a_1/j^{1/3}$. If this same process repeats, with the same $j$ value, for each of  these initial $j$ tidal disruption products when they reach $x_2= (a_2/B)^{2/3}$ (provided $x_2>1$), and so on, after $k$ such stages the original object comprises $j^k$ pieces each of size $a_1/j^{k/3}$ at distance $x_k=(a_k/B)^{2/3}$ until $x_k<1$ or $a_k<B$. This shows that, in this regime, regardless of the value of $j$, tidal fragmentation leads to arrival near the photosphere of fragments each of size close to the  critical tidal disruption size $B$ there. This means that objects of initial size $a_1\gg B$ are all reduced to a size $\sim B$ near $x=1$ by a fragmentation process lying somewhere between a large number $k$ of successive splittings, each yielding a small number $j$ of pieces, or a small number $k$ of successive splittings, each yielding a large number $j$ of pieces such that $j^{k} \sim (a_1/B)^3$.

These results indicate that, for any given value of impactor parameters - especially strength $S$ -  tidal forces ensure that fragmentation will break up any object larger than a clear lower limiting size before it reaches the photosphere. In Section 6.3 we evaluate these size limits and their consequences for the depths which impactors can reach before their explosive destruction.

\end{enumerate}

In this Section we want to determine in which of these fate domains, and at what end depth, the  destruction of any specific infaller lies as a function of its initial incident size $a_o$, for specified  values of its intrinsic properties ($\rho,{\mathcal L}, S$) and those of the star ($T_{\rm WD},M_{\rm WD}$). It is apparent from Sections 4.2 and  4.3 that, for given $a_0$, there are only two controlling parameters, $A$ and $B$.  These parameters are given by Equations (\ref{AWD}) and (\ref{BWD}) as simple products of powers of the properties of the infaller and of the star. The dominant (most wide-ranging) physical parameter factors are $T_{\rm WD}$ within $A$, whose magnitude is a measure of sublimation rate, and $S$ within $B$, whose magnitude is a measure of the material binding opposing fragmentation (a high value of $B$ indicates that fragmentation is {\textit {less} likely}).  In Section 5.2 we use the equations of Section 4 to establish a diagram of domains in the $(A,B)$ plane showing where destruction modes lie. In fact because the equations are separately linear in $A,B,a_o$, we can condense the destruction domain diagrams for all $a_o$ into a single diagram in the plane $(\alpha,\beta)$, where 

\begin{eqnarray} 
\alpha  &=& A/a_0 \nonumber\\
\beta  &=& B/a_0 
\label{defalphaandbeta}
\end{eqnarray} 

\subsection{Destruction domains in the  $(\alpha,\beta)$ plane}
We first consider in what $\alpha,\beta$ regimes fragmentation arises (i.e. in what $A,B$ regimes for a given size $a_0$).  As infall progresses, the ratio $a_{\rm frag} (x)/a_{\rm sub} (x)$ declines. In order for fragmentation to occur this ratio must reach unity or less -- and must do so at a value of $x>1$ -- i.e. outside the star  since impact destroys the body at $x=1$. In other words the two $a(x)$ functions must cross, or at least touch, with fragmentation onset at  some point $x$ with $x_2$ the largest such $x$  (since the infaller reaches $x_2$ first). The equation  $a_{\rm sub}(x)=a_{\rm frag}(x)$ for such intersection can be written
\begin{equation}
f_{\rm cross}(x) = \frac{A}{{x}^{1/2}} +Bx^{3/2} = a_0
\label{crossingpoint}
\end{equation}
where the crossing function $f_{\rm cross}(x)$ has a U-shape -- see Figure 2 -- with a minimum value $a_{\rm 0crit }(A,B)$ occurring at $x_{\rm crit}(A,B)$ where $f'_{\rm cross}(x)=0$, namely

 \begin{equation}
x_{\rm crit} =\left(\frac{A}{3B}\right)^{1/2}
\label{defxcrit}
\end{equation}

\begin{equation}
a_{\rm 0crit} = \Gamma\times A^{3/4}B^{1/4}=\Gamma a_0 \times \alpha^{3/4}\beta^{1/4}
\label{defaocrit}
\end{equation}
with 
\begin{equation}
\Gamma = \left[3^{1/4}+1/3^{3/4}\right] \approx 1.75
\label{Gamma}
\end{equation}

%%%%%%%%%%%%%%%%%%%%%%%%%%%%%%%%%% 
\begin{figure}
\centerline{
\includegraphics[width=8cm,height=9cm]{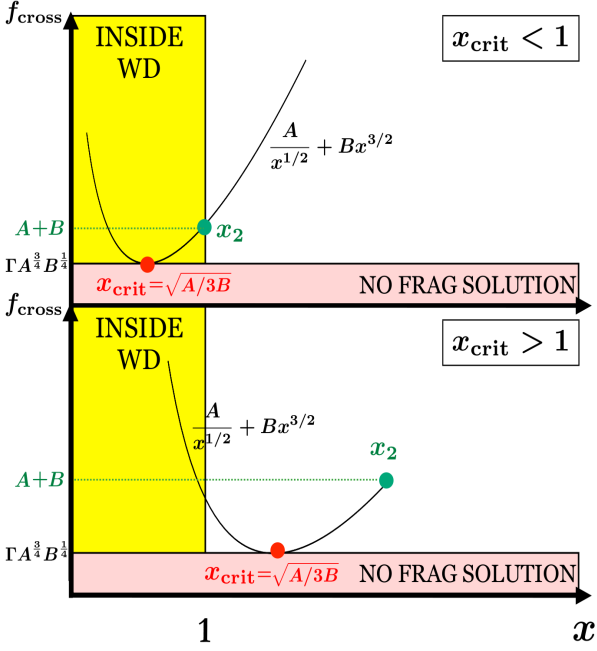}
}
\caption{
Recalling that $x \equiv r/R_{\star}$, $A$ is the sublimation parameter and $B$ is the binding parameter, presented here is a two-panel schematic of Equation (\ref{crossingpoint}), showing the form of $f_{\rm cross}(x)$ for the two cases when the minimum occurs at $x_{\rm crit}<1$ (upper panel) and $x_{\rm crit}>1$ (lower panel) and how this form influences the position $x_2$ of fragmentation onset if it occurs.  $a_0$ corresponds to a horizontal line, and fragmentation occurs if and when this line first hits the curve at a point (which may be $x_1$ or $x_2$ or neither depending on the parameters) where $x > 1$.  In the red-shaded zones where $f_{\rm cross} <a_{0{\rm crit}}=\Gamma A^{3/4}B^{1/4}$, fragmentation cannot occur for the reasons discussed in the text. 
}
\label{FigBig2}
\end{figure}
%%%%%%%%%%%%%%%%%%%%%%%%%%%%%%%%%% 

%%%%%%%%%%%%%%%%%%%%%%%%%%%%%%%%%% 
\begin{figure}
\centerline{
\includegraphics[width=8cm,height=9cm]{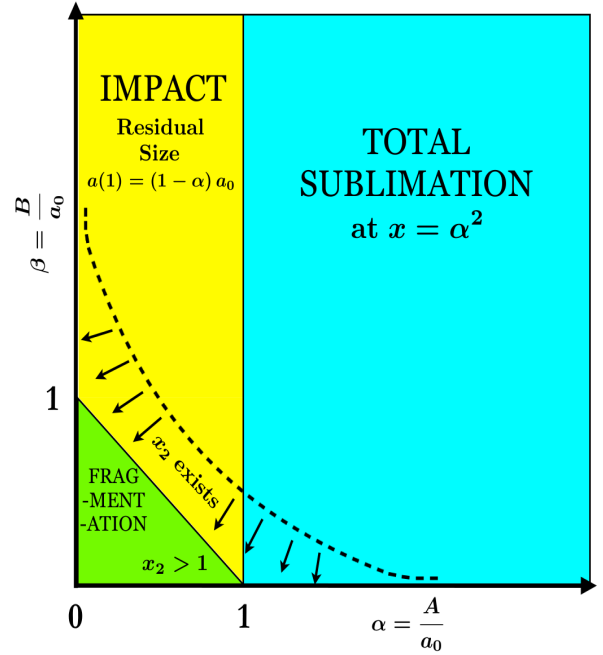}
}
\caption{
The three distinct domains of infaller destruction in the $\alpha,\beta$ plane - total sublimation, impact after partial sublimation, and fragmentation after partial sublimation. Fragmentation is restricted to the green  triangular domain in the bottom left corner. Note that $\alpha$ scales as $\propto T^4_{\rm WD}/(M_{\rm WD}\rho{\mathcal L}/a_0)$ while $\beta$ scales as $S^{1/2}/(\rho^{1/2}M_{\rm WD})$. Recall that $x \equiv r/R_{\star}$, $A$ is the sublimation parameter and $B$ is the binding parameter.
}
\label{FigBig3}
\end{figure}
%%%%%%%%%%%%%%%%%%%%%%%%%%%%%%%%%% 

Turning now to the $\alpha,\beta$ plane shown in Figure 3, using Equation (\ref{defaocrit}) we note that the first {\textit {necessary}} condition given above for fragmentation to occur is $a_0>a_{\rm 0crit}$ which can be written $\Gamma a_0 \alpha^{3/4}\beta^{1/4}<1$ or
\begin{equation}
\beta<\frac{1}{\Gamma^4\alpha^3}= \frac{27}{256\alpha^3}
\label{betaless}
\end{equation}
This  upper bound $\beta(\alpha)$ on fragmentation is shown as a dotted curve in Figure 3.
The second condition necessary for fragmentation to occur is that the solution $x_2$ of Equation (\ref{crossingpoint}) must satisfy $x_2>1$.
In terms of $\alpha$ and $\beta$, Equation (\ref{crossingpoint})  can be written $\alpha x^{-1/2}+\beta x^{3/2} = 1$. The limiting case for fragmentation to just occur is $x_2=1$ such that $\alpha+\beta=1$, or $A+B=a_o$. In other words only if 
\begin{eqnarray}
A+B&>& a_o \nonumber\\
\alpha+\beta&>&1
\label{AplusB} 
\end{eqnarray}
can fragmentation occur above the photosphere. The meaning of this is physically simple. In the absence of fragmentation an object of 
initial size $a_0$ would sublimate down to size $a_0-A$ at the photosphere ($x=1$), where the size limit for fragmentation has come down to
$Bx^{3/2} = B$. Hence, only if $a_0-A > B$ (i.e. $\alpha+\beta <1$) can fragmentation occur before impact. This upper limiting line is also shown in Figure 3, where we see that for all $x$ it lies below the first limiting curve $\beta(\alpha)$ established above (though only just below at one point). So if condition (\ref{AplusB}) is satisfied then so is condition  (\ref{betaless}) and the fragmentation domain is purely the green triangular region in the bottom left corner of the ($\alpha,\beta$) plane in Figure 3. The significance of this triangular region is that it lies at low enough $\alpha, A$ (e.g. low $T$, high $\cal L$) that sublimation does not prevent the infaller surviving far enough in to experience strong tidal gradients. At low enough $\beta, B$ (e.g. low $S$) the binding strength of material might not be strong enough to prevent fragmentation.

Having established the fragmentation domain in the  ($\alpha,\beta$) plane  it is simple to divide the rest of the plane between total sublimation prior to impact and impact prior to partial sublimation. Outside of the fragmentation domain,  an infaller will or will not lose its entire mass to sublimation according to whether $A>a_0$ or $A<a_0$  which means $\alpha>1$ or $\alpha<1$. In the first case an infaller either is totally sublimated above the photosphere at $x=\alpha^2=(A/a_0)^2$, or impacts the dense stellar ``surface'' after being sublimatively reduced in size to $a_0(1-\alpha)$ as shown by the $\alpha,\beta$ domains in Figure 3.

It only remains to determine the distance $x_2R_{\rm WD}$ at which fragmentation sets in, if at all, for a given $a_0$. There are several possible approaches to this. One is to calculate and plot the sublimated size $a(x) = a_0(1-\alpha/x^{1/2})$ for a fine grid of $x$ values commencing at $x \gg 1$ and going down to $x=1$ but stopping the plot at the first of the following conditions to be reached: (i) the total sublimation point $x_{\rm sub}=\alpha^2$ where $a(x)=0$; (ii) the fragmentation onset point $x=x_2$ where $a(x)=Bx^{3/2}=a_0\beta x^{3/2}$; (iii) impact at $x=1$ with $a=a(1)=1-\alpha$. In Figure 4 we show such plots for a set of fiducial parameters: rocky and snowy infallers (left and right panels), and cool, warm and hot WDs (top, middle and bottom rows), all for $M_{\rm WD}=0.6M_{\sun}$

%%%%%%%%%%%%%%%%%%%%%%%%%%%%%%%%%% 
\begin{figure*}
\centerline{
\includegraphics[width=8cm,height=6cm]{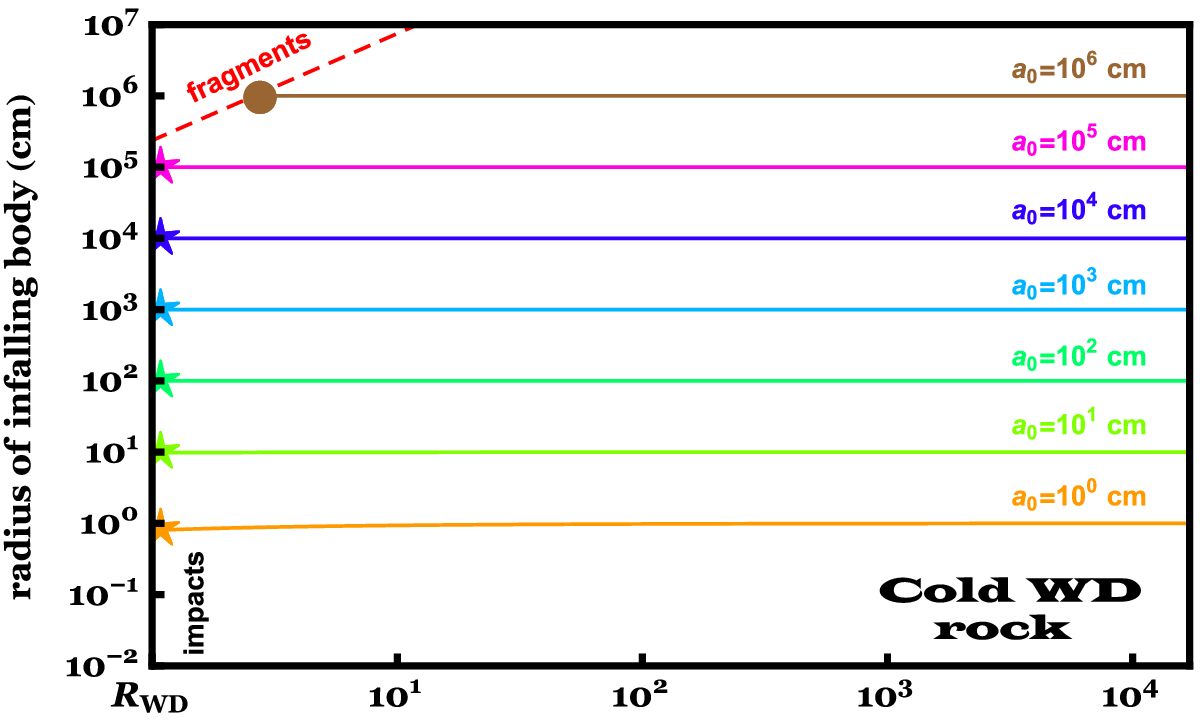}
\includegraphics[width=8cm,height=6cm]{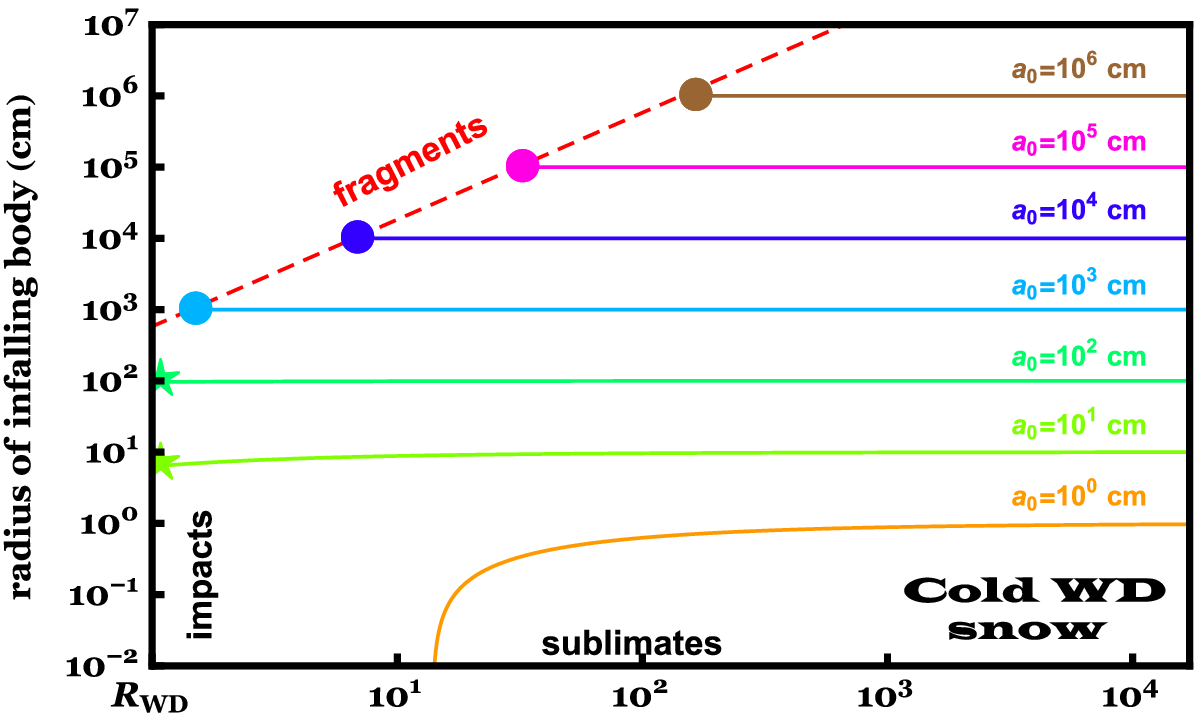}
}
\vspace{-0.35in}
\centerline{
\includegraphics[width=8cm,height=6cm]{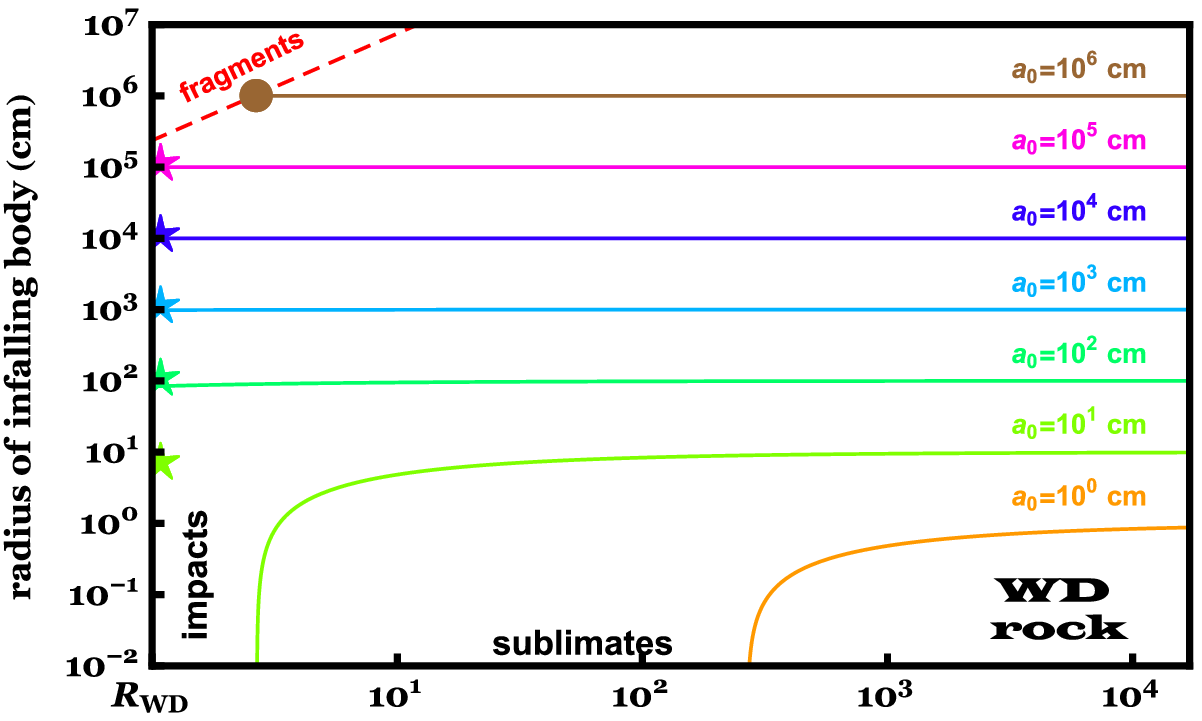}
\includegraphics[width=8cm,height=6cm]{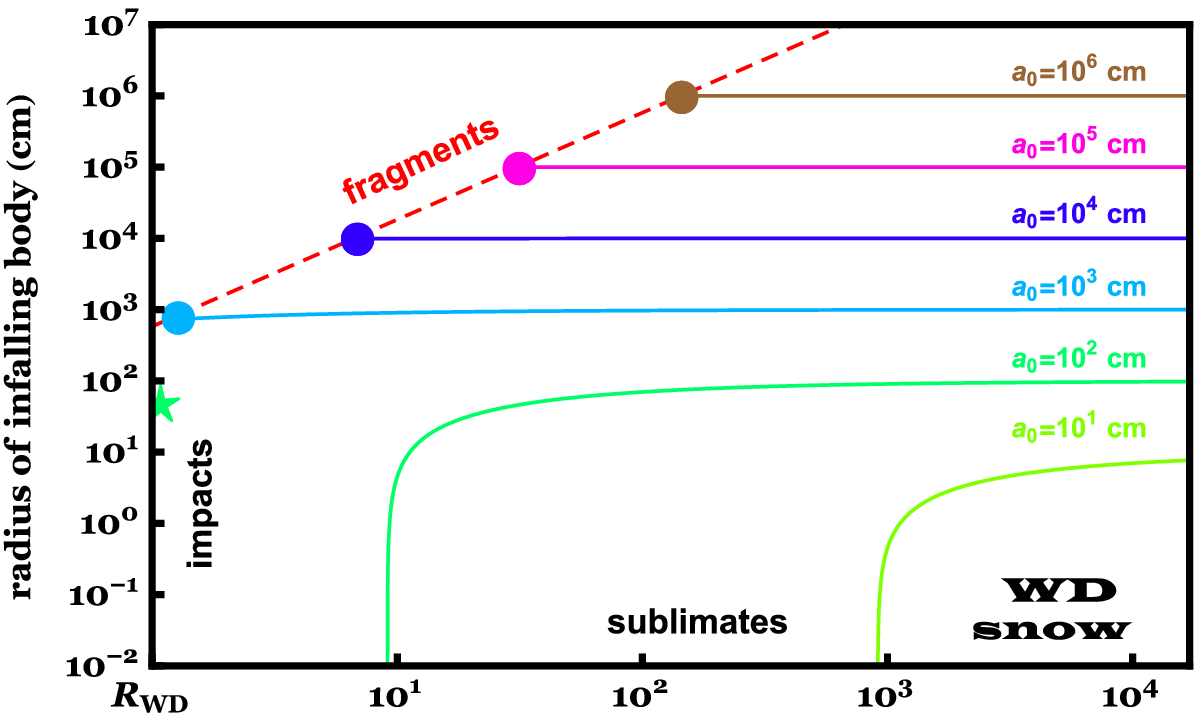}
}
\vspace{-0.35in}
\centerline{
\includegraphics[width=8cm,height=6cm]{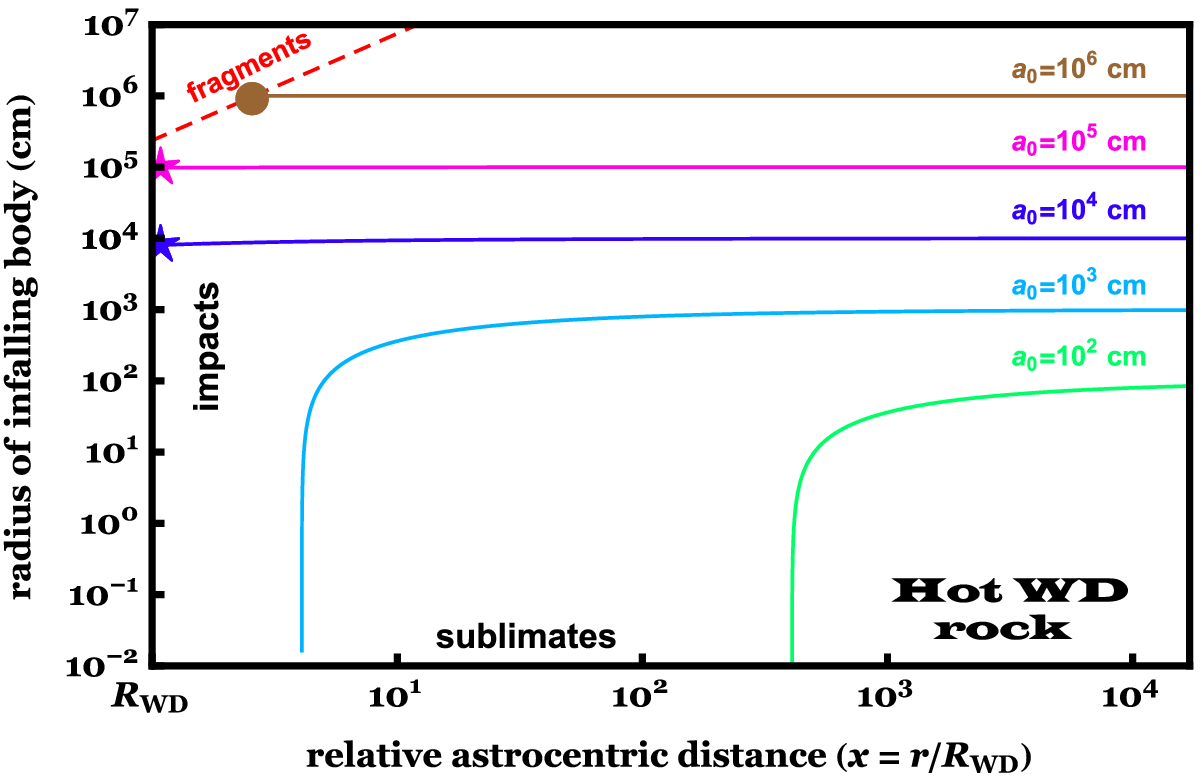}
\includegraphics[width=8cm,height=6cm]{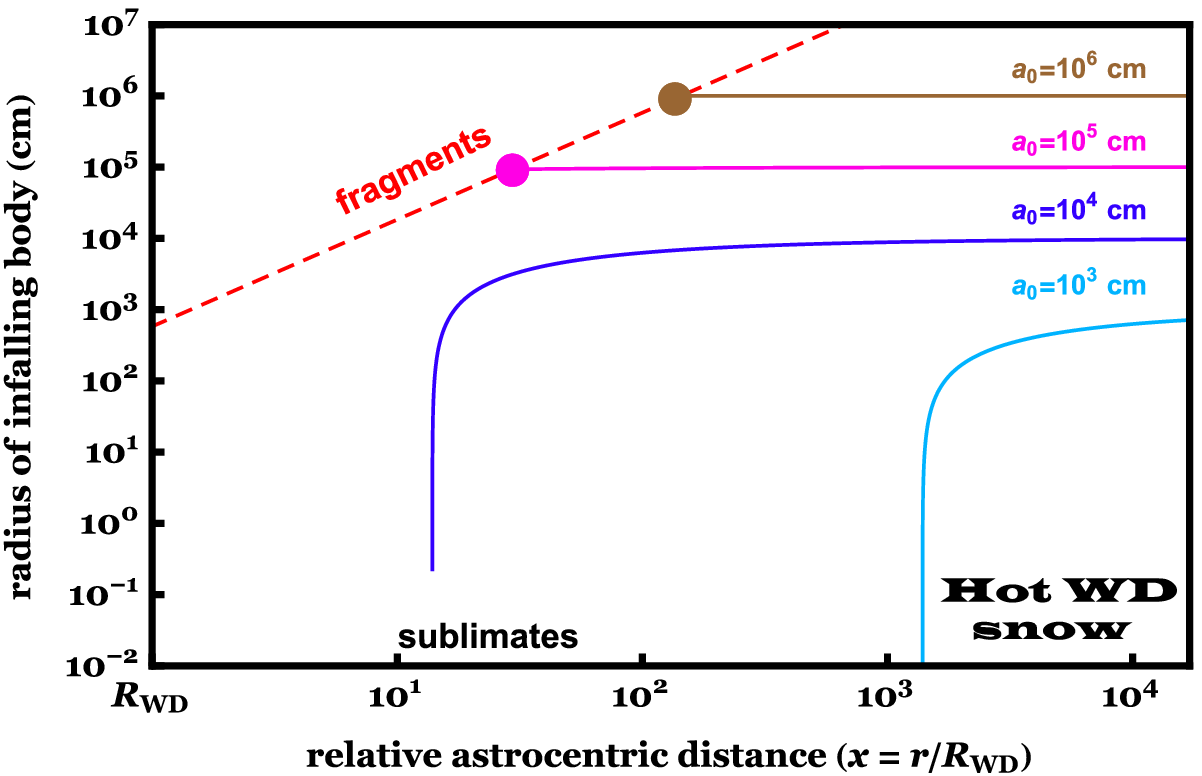}
}
\caption{
Size evolutions and fates for rocky (left panels) and snowy (right panels) bodies falling in toward a white dwarf.  Three white dwarf temperatures are sampled (5,000 K, top panels; 15,000 K, middle panels; 50,000 K, lower panels).  The initial sizes of the infallers ($a_0$) are illustrated within each plot, and range from $1$ cm to $10^6$ cm.  The three possible outcomes include (i) total sublimation (bottom axis), (ii) fragmentation (dashed red line), or (iii) impact with the WD photosphere (left axis).
}
\label{FigBig4}
\end{figure*}
%%%%%%%%%%%%%%%%%%%%%%%%%%%%%%%%%% 

If one only wants the fragmentation point value $x_2$ and not the full trajectory $a(x)$, then an alternative approach is to solve numerically the equation $\alpha/x_2^{1/2}+\beta x_2^{3/2}=1$ for $x_2(\alpha,\beta)$ over a large grid of $\alpha,\beta$ values, and use these as a look-up source of $x_2$ values or as the basis of a plot of $x_2(\alpha,\beta)$. This plot could either be a set  of smooth curves of $x_2(\alpha)$  for a series of discrete $\beta$ values, or as a plot of iso-$x$ value contours on the $(\alpha,\beta)$ plane.  Here we have adopted the look-up grid option -- shown as the matrix $\bf{X2_{\beta,\alpha}}$ -- in Table 5 of values $x_2$ for logarithmically-spaced steps in the ranges $10^{-3}\le \beta\le 1$ and $10^{-6}\le \alpha\le 1$.

Figure 5 provides, in terms of parameters $A$ and $B$, a flowchart of how to determine the fate of an infaller of size $a_0$.

\subsection{Summary}

%%%%%%%%%%%%%%%%%%%%%%%%%%%%%%%%%% 
\begin{figure}
\centerline{
\includegraphics[width=10cm,height=8cm]{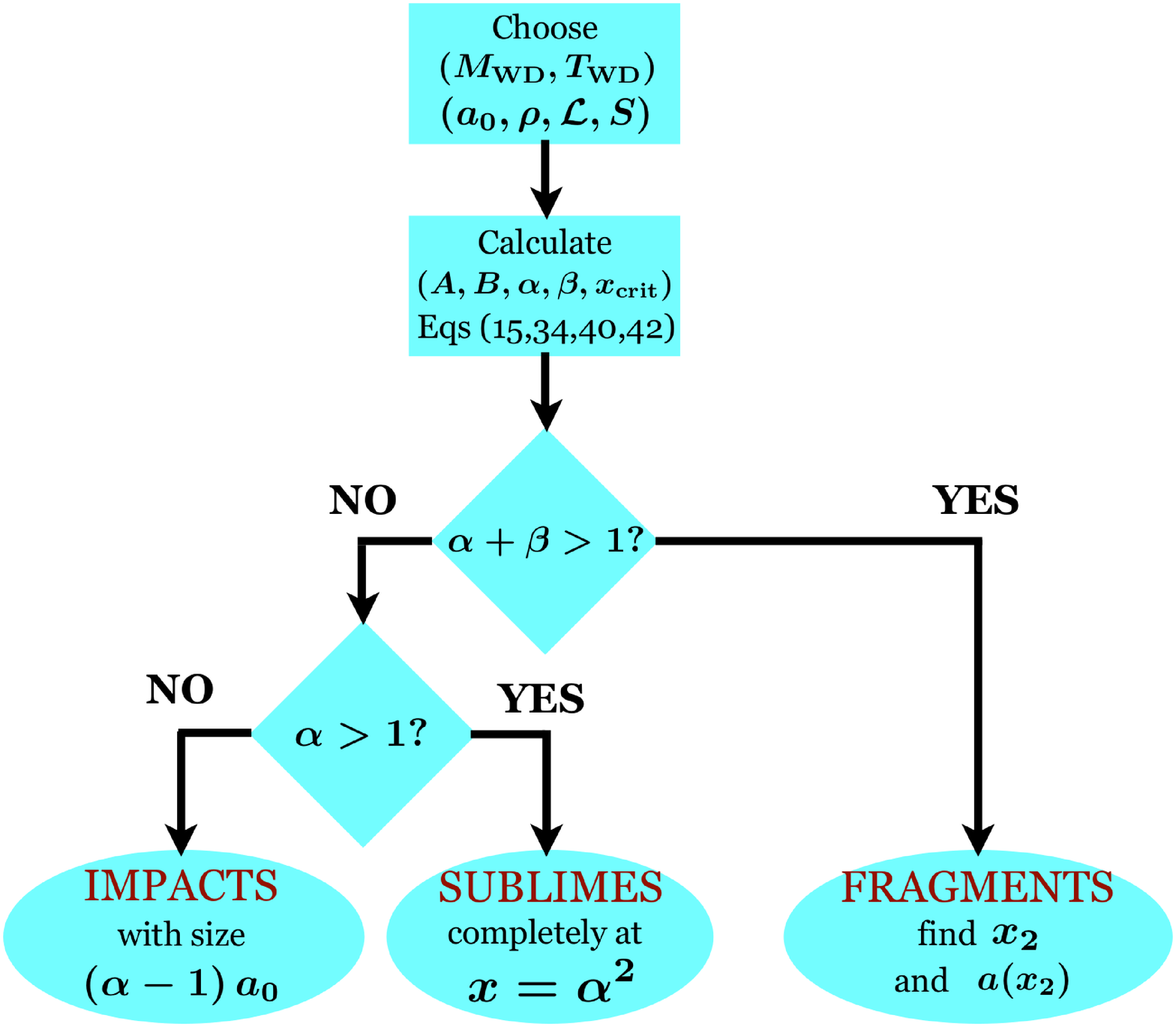}
}
\caption{Schematic flow chart of how to determine the mode and position of destruction of any infaller for any WD star starting from adopted values of the physical parameters of each.
}
\label{FigBig5}
\end{figure}
%%%%%%%%%%%%%%%%%%%%%%%%%%%%%%%%%% 

The essence of our destruction domain results  in terms of numerical values of physical parameters can best be illustrated by looking at our typical examples for parameters with our fiducial values of $\rho, {\cal L}, S$ for rock and snow, namely 
\begin {itemize}
\item {\it FRAGMENTATION} occurs if $\alpha + \beta<1\rightarrow a_0>A+B$ which, using Equations (\ref{AWDRock}), (\ref{AWDSnow}), (\ref{BWDRock}) and (\ref{BWDSnow}) for our fiducial rock and snow cases imply, respectively,

\begin{equation}
 \frac{1.3\times 10^5}{M_{\rm WD}/M_{\sun}}\left[1+1.4\times10^{-5}\left( \frac{T_{\rm WD}}{10^4 \, {\rm K}}\right)^4\right] < a_{0{\rm rock}}({\rm cm})
\label{Cond1rock}
\end{equation}
and
\begin{equation}
 \frac{3.2\times 10^2}{M_{\rm WD}/M_{\sun}}\left[1+0.10\left( \frac{T_{\rm WD}}{10^4 \, {\rm K}}\right)^4\right] < a_{0{\rm snow}}({\rm cm})
\label{Cond1snow}
\end{equation}

\item On the other hand if $ a_0<A+B$ and $a_0>A$ i.e.

\[
\frac{1.8T_4^4}{M_{\rm WD}/M_{\sun}}<  a_{0{\rm rock}}({\rm cm})
\]
\begin{equation}
\ \ \ \ \ \ \ \ \ \ \ \ \ \ \ \ <\frac{1.3\times 10^5}{M_{\rm WD}/M_{\sun}}\left[1+1.4\times10^{-5}\left( \frac{T_{\rm WD}}{10^4 \, {\rm K}}\right)^4\right] 
\label{Cond2rock}
\end{equation}
or

\

\

\[
\frac{33T_4^4}{M_{\rm WD}/M_{\sun}}  < a_{0{\rm snow}}({\rm cm})
\]
\begin{equation}
\ \ \ \ \ \ \ \ \ \ \ \ \ \ \ \ <\frac{3.2\times 10^2}{M_{\rm WD}/M_{\sun}}\left[1+0.10\left( \frac{T_{\rm WD}}{10^4 \, {\rm K}}\right)^4\right] 
\label{}
\end{equation}
then {\it IMPACT} occurs before fragmentation or complete sublimation while if $a_0<A$ (and so $ a_0<A+B$  too)  i.e.
\begin{equation}
a_{0{\rm rock}}({\rm cm}) < \frac{1.8T_4^4}{M_{\rm WD}/M_{\sun}}
\label{Cond2snow}
\end{equation}
or
\begin{equation}
a_{0{\rm snow}}({\rm cm}) < \frac{33T_4^4}{M_{\rm WD}/M_{\sun}}
\label{}
\end{equation}
then {\it SUBLIMATION} is complete before fragmentation or impact.

\end{itemize}

These results are shown graphically in Figures 6a and 6b for rock and for snow respectively, with the fiducial reference values of the various parameters given in Table 1.

%%%%%%%%%%%%%%%%%%%%%%%%%%%%%%%%%% 
\begin{figure*}
\centerline{
\includegraphics[width=9.2cm,height=7cm]{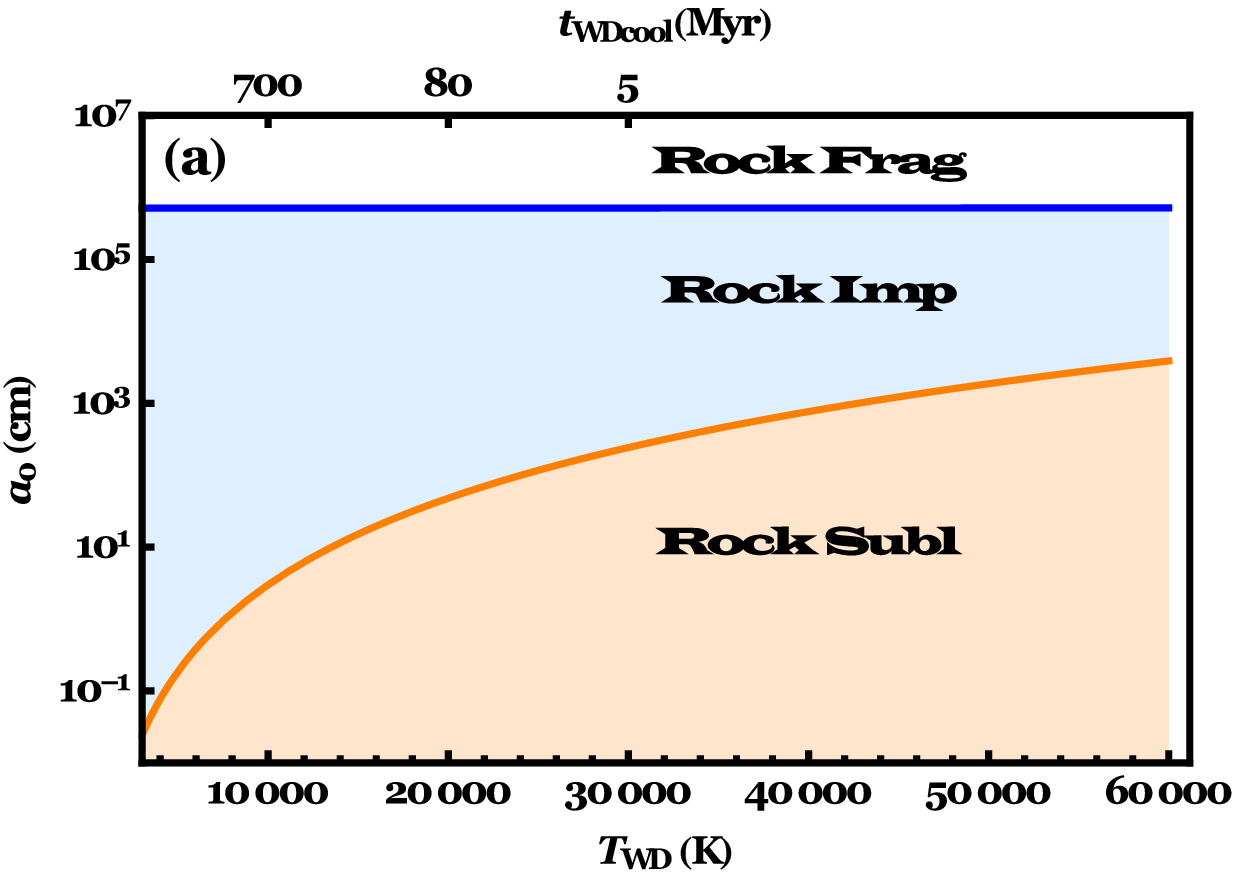}
\includegraphics[width=9.2cm,height=7cm]{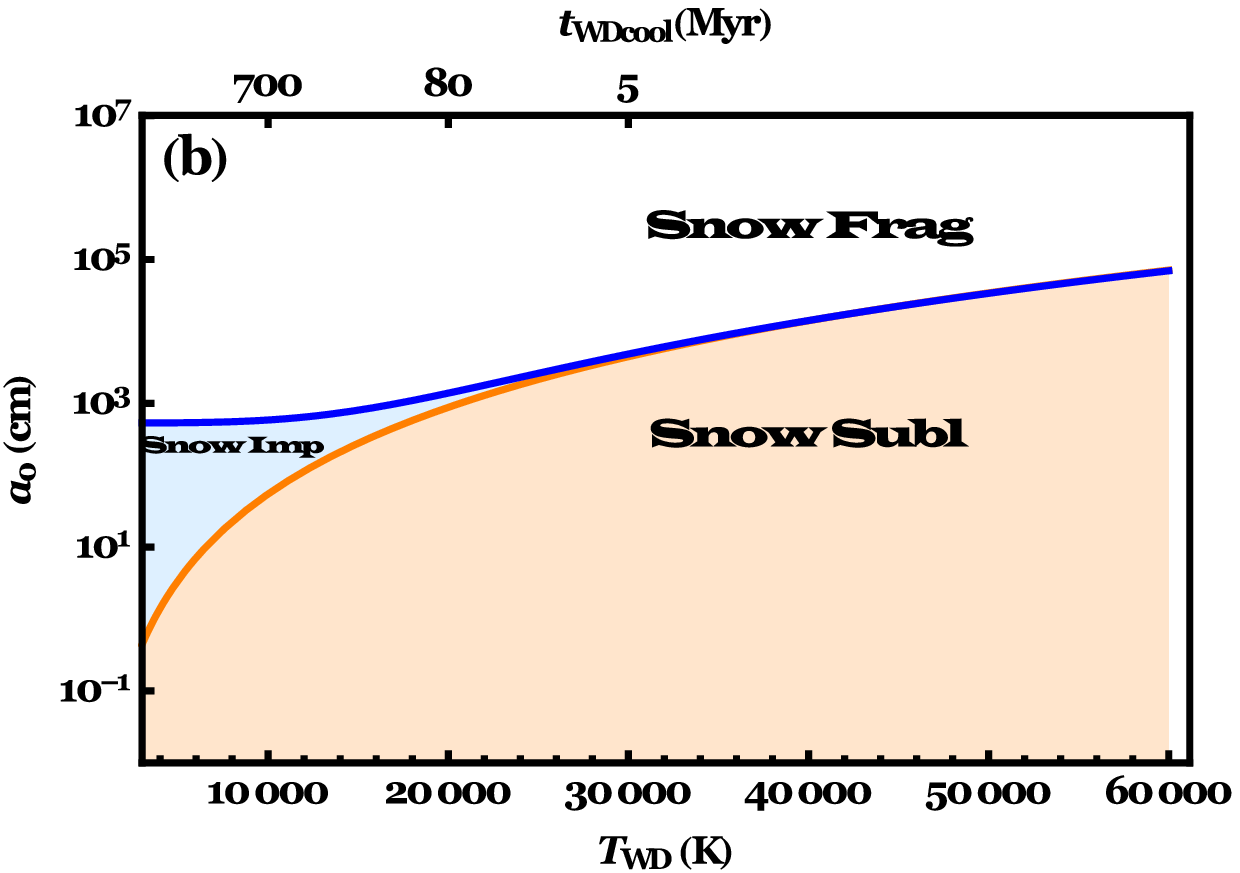}
}
\caption{Temperature dependence of destruction regimes in terms of $a_0(T_{\rm WD})$ 
for (a) rocky and (b) snowy bodies with the fiducial reference values of infaller parameters $\rho, {\cal L},S$ given in Table 1. These show that for solid rock only objects larger than several km undergo fragmentation, with impact after partial sublimation being the dominant fate in the mm - several km range for the coolest WDs, while for the hottest WDs total sublimation occurs for sizes up to around 10 m with impact occurring for larger objects up to over 1 km. For weak snowy material, for WDs of well below 30,000 K, objects over a few metres fragment and only objects below a few cm sublimate fully before fragmenting. However, above 30,000 K, total sublimation dominates for all sizes under about 1~km. Even for the coolest WDs only snowy objects in the few mm to 10 m range manage to impact before sublimating fully or fragmenting.}
\label{FigBig6}
\end{figure*}
%%%%%%%%%%%%%%%%%%%%%%%%%%%%%%%%%% 

\section{The actual `stellar impact' regime:  bow-shock ablative destruction}

\subsection{Background -  comparison with solar debris impacts}
Our description here of the destruction of bodies actually directly impacting WD photospheres is based on extension of the \cite{broetal2015} analysis of sun-impacting comets. The case of non-compact stars in general can be similarly treated by a suitable generalisation of parameters from the solar case.

Our results above show that regimes exist in which infallers (or fragments of them) can arrive near the photosphere ($r-R_{\rm WD}\ll R_{\rm WD}$) of WDs without being fragmented or fully vaporised by sublimation in starlight.  The crucial property of this region of WD atmospheres is that, because of the very high gravity $g$ and the moderate temperature $T_{\star}$ the star has a very small density scale height (Equation \ref{scaleheight}): $H\sim 10-100$ m $\preceq 10^{-2}R_{\rm WD}$. The kinetic energy flux of the incident atmospheric flow for atmospheric mass density $\rho_a$ is $\sim \rho_a(r)m_pv_o^3/2$, which increases exponentially with decreasing $r$ on distance scale $H$. The incident radiation flux causing sublimation $F_{\rm rad}\sim 6\times10^{11} (T_{\rm WD}/10^4)^4$ erg/cm$^2$/s varies very slowly (length scale $\sim R_{\rm WD}$) along the path of the infaller - roughly $\propto 1/r^2$ where $r\approx R_{\rm WD}$.  Thus, as infall proceeds, atmospheric frictional heating  very rapidly exceeds radiative. In order to assess the atmospheric density $\rho_a^{\rm subtoab}$, where the crossover occurs from radiative sublimation to atmospheric ablation, we first note that only a small fraction $C_H$ of the total  incident atmospheric bombardment flux $\rho_a(r)m_pv_o^3/2$ actually reaches the nucleus and ablates it. The remainder goes into heating the atmosphere through a stand-off bow shock which decelerates the impacting body and ablates the  nucleus by radiative, conductive and convective heat transfer. For $C_H=10^{-2}C_{H-2}$ (see below) the ablating energy flux is  $F_{\rm ab} \sim 10^{-2}C_{H-2}\rho_a(r)v_o^3/2\sim 10^{24}\rho_aC_{H-2}$. This equals the radiative  sublimation flux $F_{\rm rad}$ quoted above 
when $\rho_a \sim \rho^{\rm subtoab}_a \sim 6\times 10^{-13}$ g cm$^{-3}\times (T_{\rm  WD}/10^4)^4$ or a proton number density $n_a^{\rm subtoab}(\rm cm^{-3})\sim 3.5\times 10^{11}$ cm$^{-3}\times (T_{\rm WD}/10^4)^4$ [For further details, please see \citealt*{broetal2011}]. Thereafter the impactor is very rapidly destroyed by ablation  and deceleration within a few vertical scale heights $H$ as we describe below by analogy with \cite{broetal2015}.

 Whether or not the impactor is fully ablated before being fully decelerated or vice versa depends on the speed $v_o$ of the impactor. This speed determines the shock temperature and hence the heat transfer coefficients, according to the dimensionless parameter $X=2Q/C_Hv_0^2$, where large $X$ cases refer to the deceleration-dominated regime and small $X$ cases refer to the ablation-dominated regime. Here $Q\gg \mathcal L$ is an effective latent heat describing the total energy needed to vaporise ($\mathcal{L}$) and remove 1 g of impactor material (see \citealt*{broetal2015}).  In the case of debris impacting solar system planets, it is usually argued (e.g. \citealt*{chyetal1993,maczah1994,zahmac1994}) that the heating efficiency $C_{H}\propto 1/X$ is so small that $X$ is large and the main energy lost to the body is not ablative mass loss but simply shock heating of the atmosphere by deceleration of the body. \cite{broetal2015} have argued convincingly that, for the very high shock temperatures involved in the solar case, namely 0.4 MK for $v_0=v_{\odot} =$~620~km/s, the big increase in thermal conductivity over slow planet impact shocks means that the heating efficiency is increased ($X$ much reduced) and ablation dominates over deceleration. For WD stars ($v_{\rm WD} \approx 6200$ km/s) the shock temperature $\approx 40$ MK. This conclusion -- that ablation will dominate over deceleration -- is even stronger so below we follow \cite{broetal2015} in assuming ablative destruction dominates and in the following we use a typical fiducial value of $10^{-2}$ for the uncertain parameter $X=10^{-2}X_{-2}$.
However,  unlike \cite{broetal2015} we have to distinguish the behaviour of strong (rocky) bodies from weak (snowy) bodies. \cite{broetal2015} only discussed the latter, and in the solar context, whereas strong infalling rocks may behave differently in their response to atmospheric bow-shock ram pressure,
specifically depending on whether or not the infalling object's cross-sectional area is enhanced by compressional pancaking.

 The ram pressure of the atmosphere  impinging on an infalling body is $P_{\rm ram} =\rho v_0^2 \sim 4\times 10^{17}\rho_{\rm a}$ dyne/cm$^2$ for typical  WD $v_0$ values. A high enough  $\rho_{a}$ value can exceed the compressive strength of the body (which we take to be roughly comparable with the tensile strength $S$) and result in the body \textit{pancaking} \citep{chyetal1993,maczah1994,zahmac1994} along its path. For this to happen requires $P_{\rm ram} > S$ or $\rho_a^{\rm PramoverS} > 2\times 10^{-14}$ for $S_{\rm snow}$ and  $2\times 10^{-8}$ for $S_{\rm rock}$.  Comparing these $\rho_a^{\rm PramoverS}$ with those for  $\rho^{\rm radtoab}$ we see (see also \citealt*{broetal2015})  that for weak (e.g snowy cometary bodies) $P_{\rm ram}$ starts to drive pancaking before ablation takes over from sublimation so that pancaking is occurring through the ablation phase and the analytic bolide description \citep[e.g.][]{chyetal1993,maczah1994,zahmac1994} applies to the pancaking. On the other hand for strong rocky bodies impacting on WDs this is not the case so hard rocks will not pancake significantly in the WD impact ablation regime, except perhaps for the very hottest WDs.

An additional factor which can be important in the impact regime is the angle $\theta$ of incidence (to the vertical) of the infalling object. At distances well outside the star this angle has little effect on results and we have ignored it up till now, taking $\theta$ as always small (i.e. vertical infall and periastron distance $q=0$). However, for non zero $q$, at distance $r$, $\theta$ is given by
$\mu = \cos\theta = \sqrt{1-q/r}$ so that for $q$ near $R_\star$, $\theta$ can be a large angle with $\mu \ll 1$. The reason that small $\mu$ is so important in the impact regime is that the distance over which the atmospheric density $\rho_{\rm A}$ exponentiates \textit{along the impacting object path} is not $H$ but $H/\mu$, reducing vertical penetration by a factor $\mu$. We therefore include $\mu$ as a factor in our impact destruction depth estimates below.

We also note that, from the observability viewpoint, \cite{broetal2015} describe the result of comet-sun impacts as {\textit {cometary flares}} because of the very impulsive and local energy deposition and generation of radiation signatures like impulsive X-ray bursts and generation of helio-(astero-) seismic ripples (see \citealt*{WingandK2008} for a discussion of  WD asteroseismology). For WD impacts the impulsivity will be much greater because of the high impact speed and small scale height. Specifically the encounter speed $v_{\rm WD} = (2GM_{\rm WD}/R_{\rm WD})^{1/2}\sim 10\times v_{\odot}$, while the atmospheric density scale height $H=2kT/{\bar m}g$ is smaller by a factor $> 10^4$ for $M_{\rm WD}\approx M_{\odot}$ and $T_{\rm WD}\approx T_{\odot}$.
The other major difference from the case of sun-plunging comets is that the surface debris arrival (i.e. escape) speed is around 10 times higher, namely $\sim 6,000$ km/s for $1M_{\odot}$ and the specific impact energy  $\sim 10^4$ times higher, namely $\sim 2\times 10^{19}$ erg/g or 10 MeV/nucleon (capable of producing emissions up to gamma-ray energies). In terms of total energy,  a 1 km infaller at this speed has  a kinetic energy of $2\times10^{34}$ erg, or 10-100 times that of the largest solar flares ever observed.  The power release in impact by an infaller of size $a_0$ is ${\mathcal E} \sim a_0^3\rho v_0^2/2 >10^{32}a_0^3$ (km) erg in a timescale of a few  times $\Delta t \sim H/v_0 < 10^{-4}$ s, corresponding to an instantaneous power ${\mathcal P}\sim{\mathcal E}/\Delta t\sim 10^{36}a_0^3$ (km) erg/s which is (very briefly) $10^3$ times the bolometric luminosity $\pi R_{\rm WD}^2\sigma T_{\rm WD}^4$ of the very hottest (60,000 K) WD stars. 

So it is important to assess the observability of such impacts by their transient emissions, the discovery of which would be as exciting as the discovery of shredded planet debris orbiting near the $100R_{\rm WD}$ Roche limit via their transient absorption features \citep{vanetal2015}.

\subsection {Destruction depths for strong (rock) and weak (snow) WD impactors}
 In the following, we approximate the size of the object arriving in the lower atmosphere as being $a_0$, i.e. that of the original incident object. When an impact occurs after small fractional sublimation ($A\ll a_0$) and no fragmentation, this is quite adequate but if sublimation has significantly reduced $a_0$ to say $a=a_{\rm impact}=a_0-A$, then $a_0$ in the following expressions should be replaced by $a_{\rm impact}$.

\subsubsection{Expression for WD scale height $H$}
For all but the coolest WDs we can use the fully ionised hydrogen expression (\ref{scaleheight}) for the hydrostatic density scale height of the atmosphere which is
(see also Section 2.4)

\begin{eqnarray}
H_{\rm WD} ({\rm cm} )&=&\frac{2kT_{\rm WD}} {GM_{\rm WD} m_p/R_{\rm WD}^2}\nonumber \\&=&
\gamma^2 H_{\odot4}\frac{T_{\rm WD}} {10^4 \, {\rm K}} \left(\frac{M_{\sun}}{M_{\rm WD}}\right)^{5/3} \nonumber \\
&=& 6\times 10^3\frac{T_{\rm WD}} {10^4 \, {\rm K}} \left(\frac{M_{\sun}}{M_{\rm WD}}\right)^{5/3}
\label{HWD2}
\end{eqnarray}
where $H_{\odot4}$ is the scale height for solar surface gravity ($2.7 \times 10^4$cm/s$^2$) combined with $T=10^4$ K. We have used relationship (\ref{WDRoM}) for $R_{\rm WD}(M_{\rm WD})$.

\subsubsection{Strong/rock impactors}
In the case of vertical entry of a body with no pancaking, we define the vertical atmosphere column mass density at which ablative destruction occurs as $\Sigma_a$ (g/cm$^2$). We take this to be where the kinetic energy  $C_H\Sigma_a  v_o^2/2$ delivered per unit area (with $\Sigma_a = H \rho_a$)  equals the total energy $Q\rho a_o$ needed to drive total mass loss per unit impactor area $M/a_0^2=\rho a_0$. Allowing for non-vertical entry at angle $\theta=\cos^{-1}\mu$ it follows that most of the destruction occurs over a few vertical scale heights at a {\textit {vertical}} depth 
\begin{equation}
\Sigma_{\rm  aAblEnd}^{\rm rock}( {\rm g/cm^2})=\mu X\rho a_o\approx 3\times 10^{-2}\mu X_{-2}a_o ({\rm cm})
\label{SigmaAbRock}
\end{equation} 
where we have used the fiducial value $\rho=3$ g/cm$^3$ for rock. The corresponding hydrogen column number density is
$N_{\rm aAblEnd}^{\rm rock}({\rm cm^{-2}})=\Sigma_{\rm  aAblEnd}^{\rm rock}/m_p \sim 2\times 10^{22}\mu X_{-2}a_0 ({\rm cm})$.

The continuum optical depth (for Thomson cross section $\sigma_T \sim 7\times 10^{-25}$ cm$^2$) is $\tau = N_a\sigma_T\sim 0.01\mu X_{-2}a_o$(cm), so the destruction occurs near the continuum photosphere $\tau=1$ for vertical impact of metre (100 cm) -sized objects.

The corresponding atmospheric mass density $\rho_=\Sigma_a/H$ is

\begin{equation}
\rho_{\rm aAblEnd}^{{\rm rock}}({\rm g/cm^3})\approx\frac{5\times10^{-6} \mu X_{-2}}{T_{\rm WD}/10^4 \, {\rm K}}
\left(\frac{M_{\rm WD}} {M_\odot}\right)^{5/3}a_0 ({\rm cm})
\label{rhoaAbRock}
\end{equation} 
where we have used the above expression  for $H$ as a function of $T_{\rm WD}, M_{\rm WD}$. 
The corresponding hydrogen number density is $n_{\rm aAblEnd}=\rho_{\rm aAblEnd}/m_p$.

%%%%%%%%%%%%%%%%%%%%%%%%%%%%%%%%%% 
\begin{figure}
\centerline{
\includegraphics[width=8cm,height=7cm]{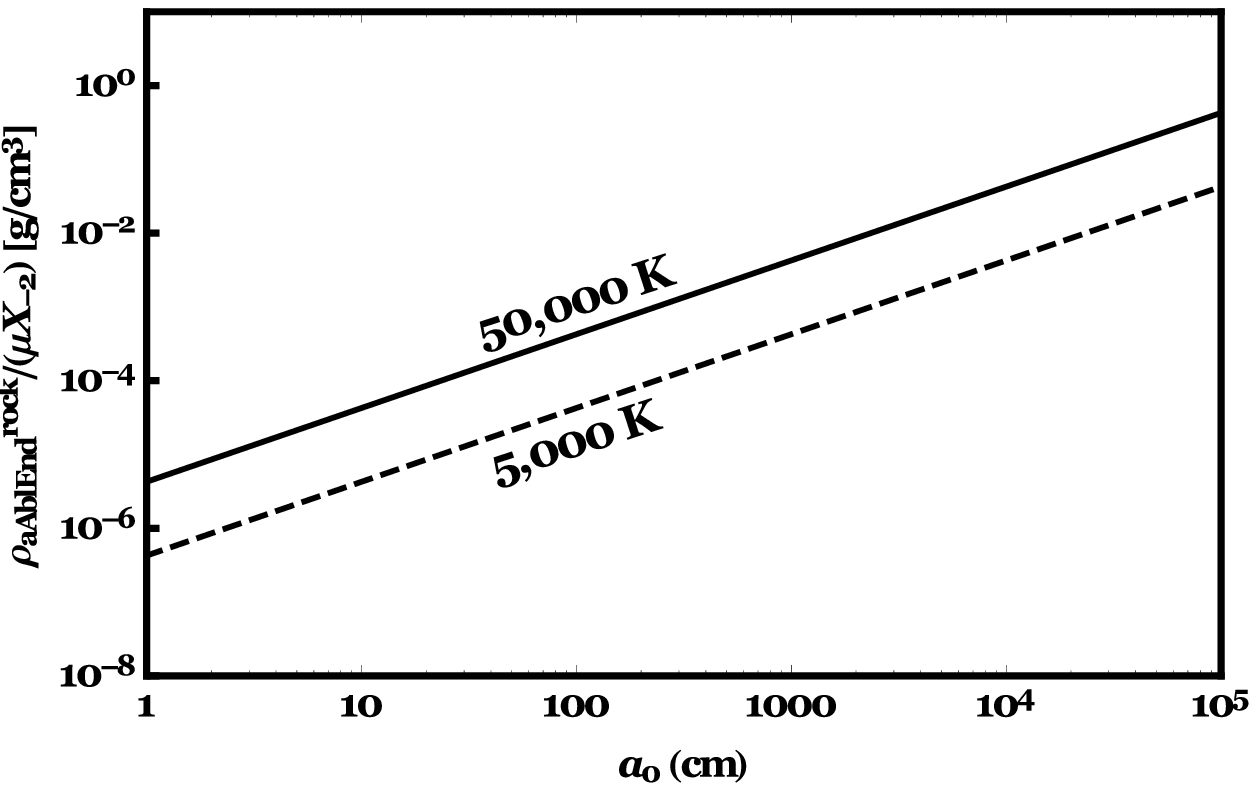}
}
\caption{The WD atmospheric mass density (g/cm$^3$) $\rho_{\rm aAblEnd}^{{\rm rock}}/\mu X_{-2}$ versus $a_0$ at the terminal impact point of infalling rocks for hot and cold fiducial values of $T_{\rm WD}$, each for two values of $M_{\rm WD}$. Strong rocks do not undergo the ram-pressure driven pancaking suffered by weaker  bodies like snow.  The $\rho$ axis has been scaled relative to the factor $\mu X_{-2}$ from Equation (\ref{rhoaAbRock}).}
\label{FigBig7}
\end{figure}
%%%%%%%%%%%%%%%%%%%%%%%%%%%%%%%%%% 

%%%%%%%%%%%%%%%%%%%%%%%%%%%%%%%%%% 
\begin{figure}
\centerline{
\includegraphics[width=8cm,height=7cm]{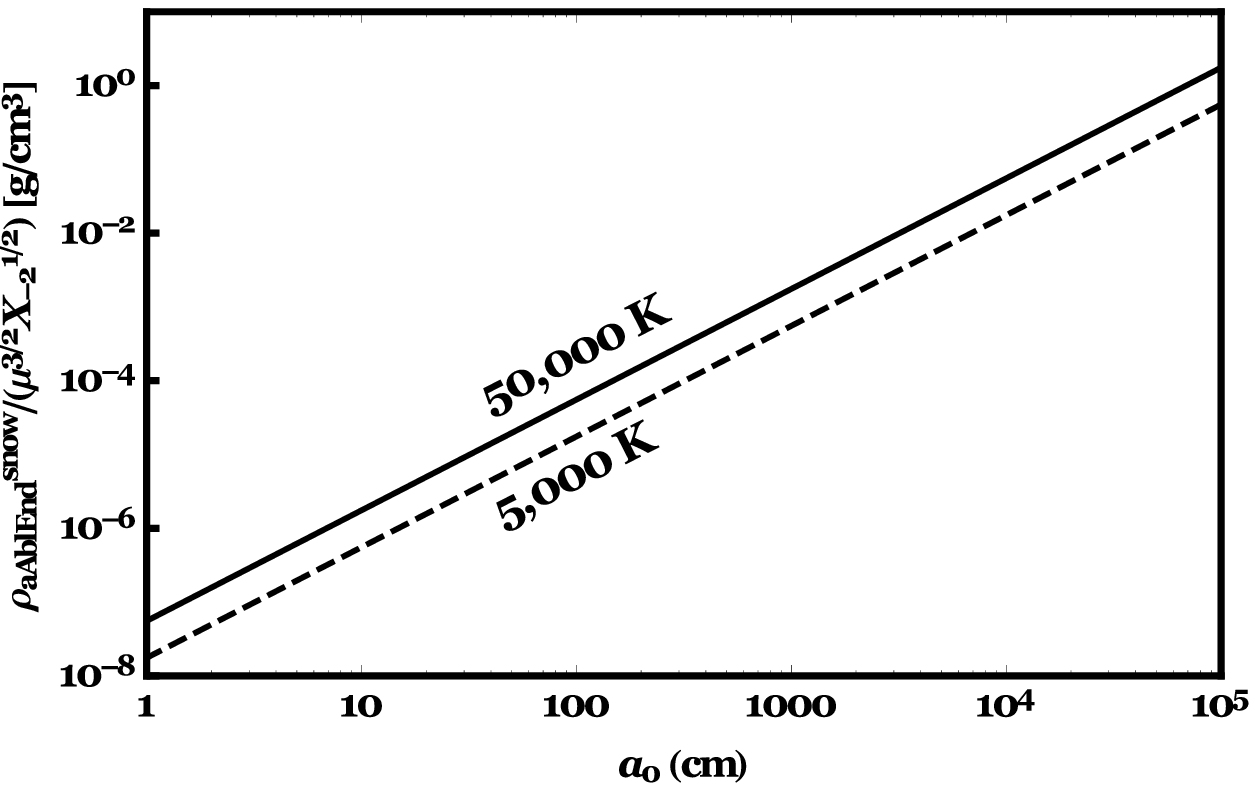}}
\caption{The WD atmospheric mass density (g/cm$^3$) $\rho_{\rm aAblEnd}^{{\rm rock}}/\mu X_{-2}$ versus $a_0$ at the terminal impact point of infalling snowy bodies for hot and cold fiducial values of $T_{\rm WD}$, each for two values of $M_{\rm WD}$. Snowy (soft) bodies undergo ram-pressure driven pancaking.  The $\rho$ axis has been scaled relative to the factor $\mu^{3/2}X_{-2}^{1/2}$ from Equation (\ref{rhoAbSnow}).}
\label{FigBig8}
\end{figure}
%%%%%%%%%%%%%%%%%%%%%%%%%%%%%%%%%% 

\subsubsection{Weak/Snow impactors}
We have argued that for WD stars and low strength impactors the ablation dominated regime of approximate analytic (ram pressure-driven) impactor pancaking radius solution discussed by e.g. \cite{chyetal1993}, \cite{maczah1994}, and \cite{zahmac1994} applies. However, we note that it depends on the condition $a_o \ll H$ and so may only be approximate for the largest bodies and coolest (smallest $H$) WDs that we consider. 
According to that pancaked solution, \cite{broetal2015} (page 7, their equation 26), the atmospheric mass column density $\Sigma_a (r)= \int _r^\infty\rho_{\rm a} dr$ (g/cm$^2$) at which ablative mass,  momentum, and energy loss peak sharply and destroy the impactor is given by

\

\

\[
\Sigma_{\rm aAblEnd}^{\rm snow} = \left(\frac{\mu^3\rho^2 a_0^3 X}{\pi H}\right)^{1/2} 
\]

\begin{equation}
= 3.6\times 10^{-4} ({\rm g/cm^2})\frac{\mu^{3/2}X_{-2}^{1/2}}{(T_{\rm WD}/10^4 \, {\rm K})^{1/2}}
\left(\frac{M_{\rm WD}}{M_{\odot}}\right)^{5/6}a_0^{3/2}({\rm cm})
\label{SigmaAbSnow}
\end{equation}
and the mass density is

\[
\rho_{\rm aAblEnd}^{\rm snow} =6\times 10^{-8} \frac{\rm g}{{\rm cm}^3}
\frac{\mu^{3/2}X_{-2}^{1/2}}{(T_{\rm WD}/10^4 \, {\rm K})^{1/2}}
\left(\frac{M_{\rm WD}}{M_{\odot}}\right)^{5/6}a_0^{3/2}({\rm cm})
\]

\begin{equation}
\label{rhoAbSnow}
\end{equation}

\noindent{}where we have again used the above expression for $H(T_{\rm WD}, M_{\rm WD})$. The dependences on $\mu, X, T,a_o$ are all more complex than in the strong rock case above because of the effects the pancaking has on the weaker material.  Estimating the optical depth $\tau$ in the same way as for the rocky body case we find that even vertically entering snowy bodies would explode above the photosphere unless they are larger than $\sim 1000$ cm.

\subsection{Effect of fragmentation size limit on maximum impactor depths}
We saw in Section 4.3 that tidal fragmentation sets upper limits to the sizes ($a(1) = B$) of rocky and snowy objects which can exist intact  at the photosphere (Equations \ref{BWDRock}-\ref{BWDSnow}). These in turn set upper limits to the maximum depths  at which explosive destruction at which rocky and snowy bodies can occur.  By inserting  these two equations into Equations (\ref{SigmaAbRock} and \ref{SigmaAbSnow}) we obtain the following  tidally limited values, using our fiducial values for rock and snow parameters

\begin{equation}
\Sigma_{\rm  aAblEndMax}^{\rm rock}( {\rm g/cm^2})\approx 4\times 10^{3}\mu X_{-2} \left(\frac{M_{\rm WD}}{M_{\odot}}\right)^{-1}
\label{SigmaAbRockMax}
\end{equation} 

and

\begin{equation}
\Sigma_{\rm aAblEndMax}^{\rm snow} = 2({\rm g/cm^2})\frac{\mu^{3/2}X_{-2}^{1/2}}{(T_{\rm WD}/10^4 \, {\rm K})^{1/2}}
\left(\frac{M_{\rm WD}}{M_{\odot}}\right)^{-2/3}
\label{SigmaAbSnowMax}
\end{equation}

The corresponding electron scattering optical depths $\tau= \Sigma \sigma_T/m_p \sim 0.4\Sigma$  are

\begin{equation}
\tau_{\rm  aAblEndMax}^{\rm rock}\approx 1600 \mu X_{-2} \left(\frac{M_{\rm WD}}{M_{\odot}}\right)^{-1}
\label{tauAbRockMax}
\end{equation} 

and 

\begin{equation}
\tau_{\rm aAblEndMax}^{\rm snow} = 0.8\frac{\mu^{3/2}X_{-2}^{1/2}}{(T_{\rm WD}/10^4 \, {\rm K})^{1/2}}
\left(\frac{M_{\rm WD}}{M_{\odot}}\right)^{-2/3}
\label{tauAbSnowMax}
\end{equation}

These equations tell us that (for $X=0.01$) the largest rocky bodies  ($10^5$ cm) arriving at the star unfragmented explode below the photosphere ($\tau =1$)  provided their entry angle cosine $\mu > 10^{-3}$ and that a vertically entering rock needs only be $\sim$ 100 cm in size to do so.
On the other hand even the largest snowy bodies  ($300$ cm) arriving at the star unfragmented and vertically will explode above the photosphere  at $\tau \sim 0.8$.

\section{Discussion and conclusions}

\subsection{Discussion}
All of the processes we have modelled above deposit the mass of individual steep fast infallers along their trajectories in the form of : sublimation products - ions, atoms, molecules and some intermingled small dust particles, too refractory and efficiently cooled to sublimate initially; in some cases, tidal fragmentation products (dust, pebbles, small boulders) with enhanced total mass sublimation rate because of the increased surface to mass ratio. Any unsublimated remains of the original object or of its fragments will ultimately either impact the photosphere or graze close by it. Apart from the direct stellar impact case, none of the processes we have modelled directly result in metallic debris deposition {\textit {on}} the WD surface ({the \textit{tabula rasa}}), where the metals are seen spectroscopically. Although the details of metal accretion are beyond the scope of the present paper, here we discuss briefly below some of the issues that require further work to answer the question of how such final deposition comes about. We also recall our conclusion (Section 4.3,  Equations \ref{BWDRock}-\ref{BWDSnow}) that tidal disruption limits the maximum size of chunks that can ever arrive near the photosphere, namely of order 1 km for rocky and 3 m for snowy bodies.

\subsubsection{Detectability and temporal signature of individual impactors}

We saw in Section 6 that the deposition of the mass and kinetic energy of impactors is extremely localised (scale of order scale height $H$) and impulsive (timescale of order $H/v_o$). This initial impact would cover a tiny fraction of the stellar disk and so be very hard to see in absorption, though its very high temperature would likely yield a briefly detectable flash of XUV emission. Furthermore, the exploding debris would spread across the stellar disk in seconds and might then be visible in absorption. We have shown that any individual impacting object is limited in size by tidal fragmentation to about 1 km if rocky and about 3 m if snowy, such objects containing masses $\sim 3\times 10^{15}$ g and $\sim 10^7$ g respectively or $\sim 10^{39}$ and $\sim 10^{31}$ nucleons respectively. Spread over the whole area $\sim 10^{19}$ cm$^2$ of a WD these correspond to column densities of $10^{20}$ and $10^{12}$ nucleons/cm$^2$ or $10^{20}/{\mathcal A}$ and $10^{12}/{\mathcal A}$ for species of atomic mass ${\mathcal A}$. Even for line transitions of moderate absorption cross section $\sim 10^{-17}$ cm$^2$ the optical depth is large $\sim 1000/{\mathcal A}$ for the largest (1 km) rocky bodies and easily detectable in high-resolution spectroscopy \citep{zucetal2007,koeetal2014} while even for the largest (3 m) snowy impactors the optical depth is $\sim 10^{-5}/{\mathcal A}$ and potentially within the reach of precision spectrometry while it lasts. The duration of detectability depends on the Type (DA or DB) of WD star involved and requires further work to assess, but along lines of argument like the following.

In the case of hydrogen-rich (DA-type) WD atmospheres, which are primarily radiative, the diffusive timescale for fresh contaminant matter to sink out of sight is only days to weeks ($10^{5-6}$ s) which would be roughly the detectability duration of a metallicity enhancing impact. For a single 1 km ($3\times 10^{15}$ g) rocky impactor the mass sinkage rate over that period is $\sim 10^{9-10}$ g/s, roughly similar to the range estimated by \cite{beretal2014} as the steady rate needed to sustain the observed level of surface contamination. Consequently to sustain the metallicity signature quasi-steadily would require arrival of 1 km rocks at intervals of days to weeks, or smaller masses more frequently. If the mean mass supply rate were delivered as bursts of 1 km objects at longer intervals, the metallicity time signature would be intermittent rather than quasi-steady. 
%For snowy bodies similar estimations can be made but, being limited to 3m in size they are much less massive by a factor of about $3\times 10^8$.

For Helium rich convective  DB-type WD stars, the situation is quite different and more complex due to their atmospheres being convective which has major effects: (i) the convective downflow will remove surface debris and spread it through the convection zone in hours. So, apart from that very brief transient, the contamination will not be visible until enough material has arrived to spread throughout the convection zone; (ii) convective upflow greatly increases the time (to Myr) for sinkage out of the convection zone. Thus the metallicity signature once established would persist for Myr after a cut off in debris supply.

\subsubsection{Near-miss star-grazing chunks} 
These are the residual post-sublimation parts of steep infallers, or of their tidally fragmented pieces,  which come close to impact but have $q>R_{\rm WD}$, and so orbit the star. We have seen in Section 4.3 that tidal forces limit the size of such star-grazers to around 1 km for rocky and 3 m for snowy debris.  The question of how these can ultimately end up as photospheric contaminants boils down to how they can shed their small angular momentum. This could occur by a variety of processes including mutual collisions, Poynting Robertson drag (most effective for the smallest pieces), or, most likely, by quite rapid sublimation. A 1 km object in a 6000 K radiation field like that at the surface of the sun or of a 6000 K WD would sublimate totally in a matter of hours. In an elliptical orbit extending from near the WD surface out to the Roche distance, that time is increased by roughly the ratio of the orbital period to the time $R_{\rm WD}/v_0$ spent near the star: a factor of $\sim$ 1000, while it is shortened by a factor of $10^4$ for very hot WDs ($T_{\rm WD} \sim 10T_\odot$). Thus the process of eccentric orbital decay and infall  of  boulders quickly becomes that of orbital decay of dust, atoms and ions.

\subsubsection{Radiation forces on dust} 
Residual material from infallers which have fully sublimated above the photosphere  ($x>1$) will likely  be in the form of molecules, atoms and ions plus some dust particles, the proportions of each depending on competing processes such as sputtering, dissociation and ionisation versus accretion, and molecular/atomic recombination.  Such small particles  will in general start in different gravitational orbits from their progenitor infalling rocks and ice but also be subject to forces additional to those discussed in Section 4 such as radiation pressure, Poynting-Robertson drag, and Yarkovsky effects, which can either aid or inhibit descent of material onto the WD surface.

\begin{itemize}
\item \textit{Radiation Pressure (RP)}:
For a particle of effective radiation cross-section $d^2$ (cm$^2$) and mass $M$ (g), the ratio of radially outward radiation pressure 
to radially inward gravitational force is (neglecting finite disc correction factors near the star)
\begin{eqnarray}
\Phi &=& \frac{R_{\rm WD}^2\sigma T_{\rm WD}^4}{GM_{\rm WD}c}\frac{d^2}{M}
     =
\frac{R_{\odot}^2\sigma T_{\rm WD}^4}
{GM_{\odot}c}\gamma^2\left(\frac{M_\odot}{M_{\rm WD}}\right)^{5/3}\frac{d^2}{M}=
\nonumber
\\
&=&
7\times10^{-8}\left(\frac{M_\odot}{M_{\rm WD}}\right)^{5/3} \frac{d^2}{M}\left(\frac{T_{\rm WD}}{10^4 \, {\rm K}}\right)^4
\label{radptograv}
\end{eqnarray}
\noindent{}where we have used the $R_{\rm WD}(M_{\rm WD})$ relations (equations \ref{WDRoM}-\ref{defgamma}) and have assumed that the dust particles are much larger than the effective wavelength of the starlight viz $\sim 400 \ {\rm nm}/(T_{\rm WD}/10^4 \ {\rm K})$.
\end{itemize}

Whether or not a particle can be prevented by radial radiation pressure from settling onto the star thus depends on the factor $T_{\rm WD}^4\times d^2/m$ which we discuss here for dust and below for ions and atoms. For pebbles and dust particles of density $\rho$  (a few  g/cm$^3$), $d^2/m \sim 1/\rho d$ (cm$^2$/g), so that in this case 
\begin{equation}
\Phi =\Phi_{\rm dust} \approx   
\frac{2\times 10^{-8}}{d({\rm cm})}  
\left(\frac{T _{\rm WD}}{10^4 \, {\rm K}}\right)^4
\left(\frac{M_\odot}{M_{\rm WD}}\right)^{5/3}
\label{Phi_{dust}}
\end{equation}
Consequently, even for WD stars which are as hot as $60,000$~K, only for infalling dust particles of residual size after radiative sublimation $d<10^{-5}$ cm (0.1 $\mu m$) would radiation pressure defeat gravity and drive particles outward.  For larger particles of the same density the radial force of radiation pressure is small compared to the inward force of gravity. However, larger particles than these are subject to the additional:

\begin{itemize}
\item {\textit {Poynting-Robertson effect  (P-R drag)}} \citep{buretal1979} of the radiation field, namely the transverse drag it exerts on the motion of orbiting matter which causes it to fall inward  even when the radial radiation pressure force on them is much less than gravity. Its importance depends on the size of the object and is very different for the WD case compared to the solar \citep{rafikov2011a,rafikov2011b,veretal2015a} because of different temperatures at different distances from the stars. For a detailed discussion of this effect and others see \cite{wyawhi1950}, \cite{buretal1979}, \cite{bonwya2010}, \cite{rafikov2011b} and \cite{veretal2015b}. However, a rough estimate of the force involved for transverse orbital speed $u$ is that it is the transverse force of the radiation field incident at aberrational angle $u/c$. Then the transverse equation of motion is

\begin{equation}
\frac{du}{dt}= -\frac{d^2}{M}\frac{u}{c} P_{\rm rad} = u\frac{d^2}{M}\frac{\sigma T_{\rm WD}^4}{c^2}
\label{PRMotionEq}
\end{equation}
which implies an orbital decay time from distance $xR_{\rm WD}$ of
\[
\tau_u\sim \frac{u}{du/dt}\sim \frac{M}{d^2}\frac{c^2x^2}{\sigma T_{\rm WD}^4}
\]
\begin{equation}
\ \ \ \ \sim 70 \, {\rm yr}\frac{\rho x^2 d(\rm cm)}{(T_{\rm WD}/10^4)^4}
\sim 6 \, {\rm day}\frac{\rho x^2 d({\rm microns})}{(T_{\rm WD}/10^4)^4}
\label{tauPRorb}
\end{equation}

This rough estimate suggests that for pebble-sized infallers, the P-R drag timescale is months to millenia, whereas for micron-sized dust, the timescale is minutes to months. The proportionalities in this estimate match previous formulations, whereas the numerical coefficient is dependent on several factors (see \citealt*{veretal2015b}), such as reflection efficiency and absorption efficiency. Another factor is eccentricity of the orbit, which necessitates the solution of coupled differential equations. Highly-eccentric orbits can generate infall timescales which differ by several orders of magnitude (see e.g. fig. 1 of \citealt*{veretal2015a}), basically because (as in the above case of sublimation) the process (PR drag) only acts effectively over a small fraction of the orbit near the star.

P-R drag actually represents the consequences of a special case of more general radiation forces. As outlined by \cite{voketal2015} for the Solar system and \cite{veretal2015b} for post-main-sequence exosystems, radiation changes both an object's spin and orbit. Spin-inducing changes are known as the YORP effect. This effect is particularly important if an object is spun up to breakup speed, an outcome which has been starkly observed in the Solar system \citep{harris1994,jacetal2014}. Orbital acceleration due to radiation is a combination of absorption, re-emission from immediate reflection, and re-emission from delayed reflection. The first two aspects together comprise RP and P-R drag, whereas the sometimes-neglected third aspect creates the Yarkovsky effect \citep{radzievskii1954,peterson1976}. \cite{veretal2015b} demonstrated that the Yarkovsky effect can induce changes which are several orders of magnitude stronger than the P-R drag, but ``turns on'' only for objects larger than pebbles. Consequently, P-R drag probably remains the most important radiation-based effect for debris within a few Roche radii of a WD.

\end{itemize}

\subsubsection{Radiation pressure effect on infalling atoms and ions} 
Except for the coolest WDs, or for cases of very high matter infall rate and density, we would expect the ionising radiation field close to the star to overwhelm recombination. Consequently most gaseous products of sublimation are likely to be in a fully or highly ionised state. For an ionised H atom (p,e pair) one can use $M\approx m_p=1.7 \times 10^{-24}$ g and $d^2$ of order the Thompson cross section $\sigma_T \sim 7\times10^{-25}$ cm$^2$, giving $ \Phi < 7\times 10^{-5}$ for any ($T_{\rm WD}, M_{\rm WD}$). Consequently, H-ions experience negligible RP force compared to gravity and should not be blown away by it.  For heavier species the same will be true unless the ionisation is partial and involves a much larger photo-absorption cross section $d^2$ \citep{chaetal1995,koeetal2014}. On the other hand, any infalling  ion experiences Lorentz and pressure forces of the WD magnetic field, though these should only re-route the inflow  along field lines rather than preventing it.

In the case of neutral atoms or low ionisation states the photo cross section $d^2$ can be much larger than $\sigma_T$  (e.g.  $\sim 10^{-15}$ cm$^2$  for H Lyman-$\alpha$) which would give  $\Phi$~$\sim$~$35(T _{\rm WD}/10^4)^4\left(\frac{M_\odot}{M_{\rm WD}}\right)^{5/3}$ so that, for all but the coolest WDs, the infall of any such highly absorbing atoms would be inhibited by radiation pressure. On the other hand, however, the fraction of  neutrals present decreases as the temperature increases. The relative importance of radiation pressure on atoms and ions will thus evolve as WDs age and cool.

\subsection{Main conclusions}
An outstanding issue in post main-sequence planetary science is identification of the dynamical origin of metallic pollutants on WD stars in the very common case where no slow-infall accretion disc exists of sufficient mass to be detectable thus far by IR excess resulting from reprocessed starlight, nor other means. Direct or very steep infall of pollutants in near parabolic orbits of very small perihelion and angular momentum may offer one possible solution since they involve a hard-to-detect fast, and hence tenuous inflow, especially if it is near isotropic. 

We have addressed, in greater depth than hitherto, the issue of what processes and parameters of the star and of the infallers govern the nature and radial distribution of deposition of such infalling debris objects as a function of their incident mass and composition and of the WD mass and temperature (cooling age). Our analysis is mainly analytic, producing simple expressions and an algorithm for easy application to modelling specific WD stars and their metallicity pollution data. Our results involve the incident size $a_0$ of the infaller and several of its intrinsic physical properties (density $\rho$, latent heat ${\cal L}$, tensile strength $S$) and on the WD mass ($M_{\rm WD}$) and effective temperature ($T_{\rm WD}$). However, for given $a_0$,  results are a function of only two length parameters ($A,B$), each of which is a simple product of powers of several of the complete set of physical parameters. $A$ is a measure of the importance of sublimation and depends most strongly on $T_{\rm WD}$ (age) while $B$ is a measure of resistance to tidal fragmentation and is mainly determined by $S$.

Our analysis applies  {\textit{inside}} the classical Roche limit where the stellar tidal gradient force exceeds the self-gravity of infalling objects. This force disrupts objects/structures of zero intrinsic tensile strength, such as sand or rubble piles, or objects of significant local strength that are permeated by surfaces of low or zero strength (cracks), as in the icy rubble-pile dusty-snowball models of cometary nuclei. For the typical densities of comets, asteroids and fragments thereof, the typical Roche limit is around a solar radius, or roughly 100 times a typical WD radius. Inside that, all objects are tidally disrupted along all of their planes of weakness into smaller pieces, whose tensile strength is more important than self-gravity. These pieces continue their infall intact -- undergoing mass loss by sublimation -- until they are completely vaporised, or impact or graze the star, or reach the point where tidal forces defeat their intrinsic strength and they fragment further.  

Our equations can be applied to obtain results for a very wide range of infaller sizes and of WD and infaller parameters. Here we have mainly shown results for four distinct regimes: cool/hot WDs, weak and volatile (snowy/comet-like) infallers, and strong and refractory (rocky/asteroid-like) infallers. We have also restricted the infaller incident size range considered to be $\sim$ cm - km.  Objects originally much below cm size are totally sublimated very far out, whereas objects much above a km are scarcer and tidally fragmented further out.

Our main findings are:
\begin{itemize}
\item Total sublimation above the photosphere befalls all small infallers across the whole WD temperature $T_{\rm {WD}}$ range, the upper threshold size rising with $T_{\rm {WD}}$ and 100$\times$ larger for rock than snow.
\item All very large objects fragment tidally regardless of $T_{\rm {WD}}$, the threshold for rock being $a_0 \succeq 10^5$ cm and for snow in the range $a_0 \succeq 10^3-3\times10^4$ cm over the full range of $T_{\rm {WD}}$.
\item No body can ever arrive at the surface of a WD with a residual size (after sublimation) larger than about 1 km for rocky material or about 3 m for  snowy material since it will be tidally disrupted there
\item A considerable range of $a_0$ avoids fragmentation and total sublimation, and impacts or grazes cold WDs, although the range narrows rapidly with increasing $T_{\rm {WD}}$, especially for snowy bodies.
\end{itemize}

Important future work would involve linking the results presented here with individual white dwarfs exhibitinging signatures of metal pollution in their atmospheres, and implementing a detailed deposition model resulting from debris infall. As detailed in Section 7.1, residual dust is subject to sputtering, dissociation and ionisation, whereas larger fragments might be influenced by Poynting-Robertson drag, the YORP effect and the Yarkovsky effect. The effort to better understand these effects may then be traced back to the archetectures and compositions of WD planetary systems which create such infall, and lead to a better understanding of planetary system evolution across all life stages.

\section*{Acknowledgments}

We thank the referee for their encouraging, very helpful and construtive comments, and for rederiving all of the equations without prompting. We also thank Pier-Emmanuel Tremblay for discussions on convective turn-over times in white dwarfs. JCB gratefully acknowledges the financial support of a Leverhulme Emeritus Fellowship EM-2012-050\textbackslash4 and of a UK STFC Consolidated Grant ST/L000741/1, and thanks K Simpson for proof-reading and helfpul science comments. BTG and DV benefited from the support of European Union ERC grant number 320964.

\begin{table*}
 \centering
 \begin{minipage}{180mm}
  \caption{Some Roman variables and parameters used in this paper.}
  \begin{tabular}{@{}llc@{}}
  \hline
   Variable & Explanation & Equation \#(s) \\
 \hline
 $A$ & Sublimation parameter (cm) & \ref{Adef}, \ref{Astargen}-\ref{AWDSnow}  \\[2pt]
 $a$ & Mean dimension of infaller & \ref{aofAandx}, \ref{aof1}, \ref{amaxfragfull}, \ref{amaxfrag}  \\[2pt]
 $a_0$ &  Initial infaller $a$ value & \\[2pt]
 $a_{0crit}$ &  $a(x)$ at which $a_{\rm sub} = a_{\rm frag}$ & \\[2pt]
 $a_{\rm frag}$ & Size for fragmentation onset & \ref{amaxfrag} \\[2pt]
 $a_{\rm sub}$ & Sublimated radius (at $x$) & \ref{aofAandx} \\[2pt]
 $B$ & Binding parameter (cm) & \ref{defB}, \ref{Bstar}-\ref{BWDSnow} \\[2pt]
 $C_H$ & Bow shock to nucleus heat transfer coefficient &  \\[2pt]
 $\mathcal{E}$ & Nucleus kinetic energy at impact & \\[2pt]
 $f_{\rm cross}$ & Function which defines $x$, if any, at fragmentation onset & \ref{crossingpoint} \\[2pt]
 $F_{\rm ab}$ & Ablative energy flux &  \\[2pt]
 $F_{\star}$ & Bolometric radiation flux at star surface &  \\[2pt]
 $F_{\rm rad}$ & Bolometric radiation flux at $r$ & \ref{Frad} \\[2pt]
 $\mathcal{F}_{\rm tot}$ & Net disruptive force & \ref{forcesum} \\[2pt]
 $\mathcal{F}_{\rm T}$ & Tidal force & \ref{} \\[2pt]
 $\mathcal{F}_{\rm G}$ & Self-gravity binding force & \ref{} \\[2pt]
 $\mathcal{F}_{\rm S}$ & Tensile strength force & \ref{forcevalues} \\[2pt]
 $G$ & Gravitational constant &  \\[2pt]
 $g_{\star}$ & Stellar surface gravity & \ref{gravity} \\[2pt]
 $H$ & Density scale height & \ref{scaleheight} \\[2pt]
 $k$ & Boltzmann constant  &  \\[2pt]
 $L_{\star}$ & Bolometric stellar luminosity & \\[2pt]
 $\mathcal{L}$ & Latent heat of ``vapourisation'' & \\[2pt]
 $M$ & Infaller mass & \ref{Mass(r)}, \ref{MofAandx} \\[2pt]
 $M_{\star}$ & Stellar mass & \\[2pt]
 $M_{0}$ & Initial stellar mass & \\[2pt]
 $m_{\rm p}$ & Proton mass & \\[2pt]
 $n_{\rm a}^{\rm subtoab}$ & Stellar atmosphere hydrogen number density & \\[2pt]
 $\mathcal{P}$ & Instantaneous power of energy released in impact & \\[2pt]
 $P_{\rm ram}$ & Ram pressure of atmosphere impinging on infaller & \\[2pt]
 $P_{\rm ramsub}$ & Ram pressure of sublimating mass outflow & \\[2pt]
 $q$ & Infaller periastron distance & \\[2pt]
 $Q$ & Total vapourisation energy for ablation & \\[2pt]
 $R$ & Stellar radius & \ref{WDRoM} \\[2pt]
 $\mathcal{R}$ & Ratio of self-gravity to material strength & \ref{AppEqn1a}, \ref{AppEqn1b}, \ref{AppEqn1} \\[2pt]
 $r$ & Astrocentric distance of infaller &  \\[2pt]
 $S$ & Infaller's tensile strength &  \\[2pt]
 $\Delta t$ & Timescale of impact energy deposition &  \\[2pt]
 $T$ & Stellar effective temperature &  \\[2pt]
 $v_{\star}$ & Escape speed from stellar surface &  \\[2pt]
 $v(r)$ & Infaller speed & \ref{vofr} \\[2pt]
 $X$ & $\equiv \frac{2Q}{C_{\rm H} v_{\star}^2}$ & \\[2pt]
 $x$ & $\equiv \frac{r}{R_{\star}}$ & \\[2pt]
 $x_1, x_2$ & The real solutions of Eq. (\ref{crossingpoint}), if any  & \\[2pt]
 $x_{\rm crit}$ & The $x$ value at which $a_{\rm sub} = a_{\rm frag}$ & \ref{defxcrit} \\[2pt]
 $x_{\rm sub}$ & The $x$ value where total sublimation occurs for a size $a_{0}$ &  \\[2pt]
 $x_{\rm Roche}$ & The $x$ value at the Roche radius & \ref{Rochex}  \\[2pt]
 $X_{-2}$ &  $=100X$  &  \\[2pt]
\hline
\end{tabular}
\end{minipage}
\end{table*}

\begin{table*}
 \centering
 \begin{minipage}{180mm}
  \caption{Some Greek variables and parameters used in this paper.}
  \begin{tabular}{@{}llc@{}}
  \hline
   Variable & Explanation & Equation \#(s) \\
 \hline
 $\alpha$ & $\equiv \frac{A}{a_0}$ &  \ref{defalphaandbeta} \\[2pt]
 $\beta$ & $\equiv \frac{B}{a_0}$ &  \ref{defalphaandbeta} \\[2pt]
 $\Gamma$ & $\equiv 3^{1/4} + 3^{-3/4}$ &  \ref{Gamma} \\[2pt]
 $\gamma$ & $\equiv 10^{-2}$ (WD radius factor) &  \ref{defgamma} \\[2pt]
 $\theta$ & Impactor entry angle to vertical &   \\[2pt]
 $\mu$ & $\equiv \cos{\theta}$ &   \\[2pt]
 $\rho$ & Infaller mass density &  \\[2pt]
 $\rho_{\rm a}$ & Atmospheric mass density & \ref{rhoaAbRock}, \ref{rhoAbSnow} \\[2pt]
 $\rho_{\rm a}^{\rm subtoab}$ & Atmospheric mass density crossover point from sublimation to ablation-dominated &  \\[2pt]
 $\sigma$ & Stefan-Boltzmann constant &   \\[2pt]
 $\Sigma$ & Vertical atmospheric column mass density & \ref{SigmaAbRock}, \ref{SigmaAbSnow}   \\[2pt]

\hline
\end{tabular}
\end{minipage}
\end{table*}

\begin{table*}
 \centering
 \begin{minipage}{180mm}
  \caption{$x_2$ values corresponding to ($\alpha$,$\beta$) pairs, where $\alpha$ is on $y$-axis
and $\beta$ is on the $x$-axis.}
  \begin{tabular}{c cccc ccccc c}
  \hline
            &  $10^{-3.0}$ & $10^{-2.7}$  &  $10^{-2.3}$ &  $10^{-2.0}$ & $10^{-1.7}$  &  $10^{-1.3}$ &  $10^{-1.0}$ &  $10^{-0.7}$  &  $10^{-0.3}$ & $10^{0.0}$ \\
\hline
 $10^{-6.0}$ &    100      &     63      &     34      &     22      &    14       &     7.4     &    4.6     &     2.8      &     1.7     &   1.0     \\
% $10^{-5.7}$ &           &             &             &             &             &             &            &              &             &  \\
% $10^{-5.3}$ &             &             &             &             &             &             &            &              &             &  \\
% $10^{-5.0}$ &             &             &             &             &             &             &            &              &             &  \\
% $10^{-4.7}$ &             &             &             &             &             &             &            &              &             &  \\
% $10^{-4.3}$ &             &             &             &             &             &             &            &              &             &  \\
 $10^{-4.0}$ &    100      &     63      &     34      &     22      &    14       &     7.4     &    4.6     &     2.8      &     1.7     &   NA     \\
% $10^{-3.7}$ &             &             &             &             &             &             &            &              &             &  \\
% $10^{-3.3}$ &             &             &             &             &             &             &            &              &             &  \\
% $10^{-3.0}$ &             &             &             &             &             &             &            &              &             &  \\
% $10^{-2.7}$ &             &             &             &             &             &             &            &              &             &  \\
% $10^{-2.3}$ &             &             &             &             &             &             &            &              &             &  \\
 $10^{-2.0}$ &    100      &     63      &     34      &     22      &    14       &     7.3     &    4.6     &     2.8      &     1.7     &   NA     \\
% $10^{-1.7}$ &             &             &             &             &             &             &            &              &             &  \\
% $10^{-1.3}$ &             &             &             &             &             &             &            &              &             &  \\
 $10^{-1.0}$ &     99      &     63      &     34      &     21      &    13       &     7.2     &    4.5     &     2.7      &     1.6     &   NA     \\
 $10^{-0.7}$ &     99      &     62      &     33      &     21      &    13       &     7.0     &    4.3     &     2.5      &     1.5     &   NA \\
 $10^{-0.3}$ &     97      &     61      &     32      &     20      &    12       &     6.4     &    3.9     &     2.2      &     1.1     &   NA  \\
 $10^{0.0}$ &     93      &     57      &     30      &     18      &    11       &     4.9     &    2.2     &     NA       &      NA     &   NA \\
\hline
\end{tabular}

\end{minipage}
\end{table*}

\label{lastpage}

\end{document}